\documentclass[12pt]{article}

\usepackage{amsmath,amssymb}
\usepackage{authblk}
\usepackage{enumitem}  
\usepackage[linktocpage]{hyperref}  
\usepackage[usenames, dvipsnames]{color} 
\usepackage{graphicx}  

\usepackage{xcolor}

\usepackage{latexsym}
\usepackage{amssymb}
\usepackage{amsmath}
\usepackage{graphicx}
\usepackage{enumitem}
\usepackage{slashed}
\usepackage{hyperref} 
\usepackage{endnotes}

\newcommand\mathC{\mkern1mu\raise2.2pt\hbox{$\scriptscriptstyle|$}
        {\mkern-7mu\rm C}}              

\def\be{\begin{equation}}
\def\ee{\end{equation}}
\def\bear{\begin{eqnarray}}
\def\eear{\end{eqnarray}}

\def\a{\alpha}

\def\t{\tau}

\def\a{\alpha}

\def\t{\tau}

\newcommand{\sm}[1]{\mbox{\scriptsize #1}}
\newcommand{\tn}[1]{\mbox{\tiny #1}}
\renewcommand{\@}[1]{\sqrt{#1}}
\renewcommand{\le}[1]{\label{#1}\end{eqnarray}}
\newcommand{\bea}{\begin{eqnarray}}
\newcommand{\eea}{\end{eqnarray}}

\newcommand{\eq}[1]{(\ref{#1})}

\def\ffract#1#2{\raise .35 em\hbox{$\scriptstyle#1$}\kern-.25em/
\kern-.2em\lower .22 em \hbox{$\scriptstyle#2$}}

\begin{document}

\pagestyle{empty}
\numberwithin{equation}{section}

\centerline{{\Large\bf  Emergence and Correspondence for}} 
\vskip.25truecm
\centerline{{\Large\bf String Theory Black Holes}}
\vskip.95truecm

\begin{center}
{\large Jeroen van Dongen,$^{1,2}$ Sebastian De Haro,$^{2,3,4,5}$}
\vskip.20cm
{\large Manus Visser,$^1$ and Jeremy Butterfield$^3$}\\
\vskip .75truecm
$^1${\it Institute for Theoretical Physics, University of Amsterdam}\\
$^2${\it Vossius Center for the History of Humanities and Sciences, University of Amsterdam}
$^3${\it Trinity College, Cambridge}\\
$^4${\it Department of History and Philosophy of Science, University of Cambridge}\\
$^5${\it Black Hole Initiative, Harvard University}\\

\vskip 1cm
\today
\\
\end{center}
\vskip 1cm

\begin{center}
\textbf{\bf Abstract}
\end{center}
This is one of a pair of papers that give a historical-\emph{cum}-philosophical  analysis of the endeavour to understand black hole entropy as a statistical mechanical entropy obtained by counting string-theoretic microstates. Both papers focus on Andrew Strominger and Cumrun Vafa's ground-breaking 1996 calculation, which analysed the black hole in terms of D-branes. 
The first paper gives a conceptual analysis of the Strominger-Vafa argument, and of several research efforts that it engendered. In this paper, we assess whether the black hole should be considered as emergent from the D-brane system, particularly in light of the role that duality plays in the argument. We further identify uses of the quantum-to-classical correspondence principle in string theory discussions of black holes, and compare these to the heuristics of earlier efforts in theory construction, in particular those of the old quantum theory.

\newpage
\pagestyle{plain}
 \tableofcontents

\newpage

\section{Introduction}
\label{sec:intro}

In 1996, Andrew Strominger and Cumrun Vafa offered the first microscopic calculation of black hole entropy within string theory: namely in terms of D-branes.\footnote{Strominger and Vafa (1996).} This calculation quickly gained wide acceptance, and was soon seen as one of string theory's main successes.\footnote{As an indication of the success of the Strominger-Vafa result, note that string theory critic Lee Smolin (2006:~p.~138) called it ``perhaps the greatest accomplishment of the second superstring revolution," while string theorist Thomas Banks recalled a ``flurry of activity" in its wake (2005:~p.~325). } This was due not only to the result itself---the counting of black hole microstates had been a research goal for some two decades---but also because of 
its role in later developments and its larger
implications. 
It suggested to many that black hole evolution should be a unitary process, thus influencing debates on the information paradox.  And it played a key role in developments leading up to the celebrated AdS/CFT duality.

The aim of this paper and its companion\footnote{See the article `Conceptual Analysis of Black Hole Entropy in String Theory'. For related work on the larger subject of the information paradox, see Jeroen van Dongen and Sebastian De Haro, `History and Philosophy of the Black Hole Information Paradox' (Forthcoming).} is to give a historical-philosophical analysis of Strominger and Vafa's argument, and outline its contemporary role in string theory and debates on black hole entropy.  In the first paper, we gave a conceptual analysis of the Strominger-Vafa argument, and of several research efforts that it engendered.  
Here, we address three key questions that arise from that analysis. They are, roughly speaking, as follows: 
\\
\\
(i) Is the black hole the very same physical system as the D-brane system to which it is compared? \\
(ii) Is the black hole emergent from the D-brane system? \\
(iii) Does the correspondence principle, as used in the development of quantum theory, illuminate the Strominger-Vafa argument: or more generally, the microscopic basis of black hole entropy? 
\\
\\
We first briefly review our analysis of the Strominger-Vafa argument (Section 2). Then in Section 3, we ask: Is the black hole the same physical system as the D-brane system?
What states are counted? Those of a black hole, or those of some quantum system that is in fact different from the black hole?  The question is prompted by one of the main engines of the argument: a duality between open and closed strings.  We note that for the black hole that Strominger and Vafa consider, there is {\em not}, so far as we now know, an exact duality between open and closed strings---more about this in Section \ref{ocdual}.

The question of the relation between the black hole and the D-brane system, and the assessment of this duality, raise the question whether the black hole should be considered as \emph{emergent} from the D-brane system. We address this question in Section 4. Here, we distinguish different relations that might obtain between emergence and duality. In all scenarios considered, the relation between the black hole and the D-brane system indeed appears best captured by a notion of emergence, though that relation will be of a different nature in the various cases.

Emergence links the classical black hole to the quantum system. Traditionally, and heuristically in efforts at theory construction, that relation has often been studied as an example of \emph{correspondence}. 
So in Section \ref{corrp}, we ask what comparative lessons we might learn from Niels Bohr's invocation of a correspondence principle as a heuristic in developing the old quantum theory. We will argue that, once the principle is suitably clarified, it offers  fruitful analogies with the Strominger-Vafa argument and, more generally, with current efforts to construct a quantum theory of gravity. Section \ref{conclsn} concludes.

\section{Counting Black Hole Microstates in String Theory}\label{SVarg}

We here   give the main outline of the   argument by which Andrew Strominger and Cumrun Vafa claimed to have counted the microstates of a certain type of black hole.\footnote{For a detailed treatment and introduction to the concepts, see our companion article `Conceptual Analysis of Black Hole Entropy in String Theory', especially Sections 2 and 3.} 
The argument relates two distinct systems: a classical black hole in five-dimensional supergravity (i.e.~general relativity with specific matter fields) and a configuration of D-branes, using a series of conjectured dualities in string theory. 
At low values of the coupling, the world-volume quantum field theory of the D-branes offers a microscopic, quantum description of the configuration of D-branes. It consists of two types of D-branes, namely one-dimensional D1-branes and five-dimensional D5-branes, intersecting along a common circle.\footnote{ As we noted in Section 3.2.2 of the companion paper, D3-branes are also allowed: but we discuss here the simplest version of the set-up, with no D3-branes.} The number of D-branes is $N_1$ and $N_5$, respectively. The D-branes share the same values of mass, angular momentum and electric charge as a macroscopic black hole; these are computed when the coupling is large. There are two fields sourcing electric charge and the electric charges are denoted by $Q_{\tn F}$ and $Q_{\tn H}$, where the charges correspond to the number of branes $N_1,N_5$. 
Strominger and Vafa showed that the microscopic entropy of the D-brane configuration matches the Bekenstein-Hawking entropy for the corresponding supergravity black hole, i.e., yields the same number as the horizon area of the black hole in Planckian units.\footnote{Cf. Bekenstein (1972, 1973) and Hawking (1975).}

The two quantities (the microscopic, Boltzmannian entropy calculated from the D-brane counting, and the Bekenstein-Hawking area-entropy) are calculated in different regimes of parameters. The supersymmetry of both solutions ensures that the value of the entropy does not change from one regime to the other, for suitable changes of the couplings: `adiabatic' changes.

The calculation by Strominger and Vafa had been made possible because, in late 1995, Joseph Polchinski had shown that quantum D-branes offered the possibility to link up field theory accounts with gravitational $p$-branes.
A $p$-brane is a compactification of a $p$-dimensional black hole-like solution of ten-dimensional classical supergravity (so that a 0-brane is a black hole, a 1-brane is a black string etc.) and supergravity is a low-energy limit of closed string theory. Earlier in the year Edward Witten had argued that particular black hole states in supergravity could be the classical counterparts of certain supersymmetric superstring states known as `BPS' states.\footnote{Witten (1996a).} 
Polchinski's suggestion was that his D-branes be identified with the quantum states corresponding to Witten's supergravity solutions.

For the $p$-branes, one could calculate horizon areas; for the D-branes, state counting was possible. The numbers for entropy that could thus be obtained then needed to match up. A way to fulfill that would be to find a way to link the entropies as \emph{adiabatic invariants}:
for what D-brane system is the entropy invariant when the coupling is gradually turned up (i.e.~an adiabatic invariant), and what would its corresponding $p$-brane system look like? This is the central research problem that Strominger and Vafa confronted.

They identified the systems as particular D-brane and extremal $p$-brane solutions of type~II string theory compactified on the five-dimensional internal manifold $\mbox{K3} \times S^1$, considered in appropriate regimes.\footnote{Here $S^1$  denotes a circle and $\mbox{K3}$ is a  4-dimensional compact manifold, called a `Calabi-Yau' manifold.} BPS states and extremal black holes both have a mass that is exactly equal to the charge (in appropriate units); it was therefore natural to identify the two.
Strominger and Vafa explicitly referred to the adiabatic invariance of these solutions: for the extremal black $p$-brane solution, they    pointed out that when the asymptotic value of the dilaton (which fixes the coupling constant) ``is adiabatically changed, the near-horizon geometry is unaltered.''\footnote{Strominger and Vafa (1996: ~p.~101).} In  other words, if one moves away from the supergravity regime by turning down the value of the  coupling constant, viz.~the dilaton field, the expression for the area of the black hole, and thus its Bekenstein-Hawking entropy, does not change. This was a known property of such extremal black holes, as e.g.~Strominger had argued with two co-authors only a few months earlier.\footnote{See Ferrara, Kallosh and Strominger (1995); for related work, see also Ferrara and Kallosh (1996) and Strominger (1996).} 

As the coupling goes down, the system might not even be rightly interpreted any more as a black hole or as a $p$-brane with a horizon; but still its entropy would retain the same value. Strominger and Vafa stated, while quoting a result by Gibbons and Townsend (1993), that ``intuitively" this property could be attributed to topological invariance since black holes  
can be viewed ``as solitons which interpolate between maximally symmetric vacua at infinity and the horizon.''\footnote{Strominger and Vafa (1996:~p.~101). The role of BPS states as states of solitons that interpolate between different vacua had already been recognised, in the context of quantum field theory, by Cecotti and Vafa (1993:~p.~633).} Topological invariance also played a role in the D-brane calculation, as the degeneracy of the BPS states ``is a topological quantity related to the elliptic genus,'' and was also expected to remain invariant as the coupling increases.\footnote{Strominger and Vafa (1996:~p.~103).} The `elliptic genus' is a quantum field theory partition function in which fermionic states contribute with a minus sign, such that bosons and fermions cancel each other out in all states except for the BPS states. Thus, the elliptic genus is a partition function for the BPS states, and it is invariant under continuous variations of the parameters.\footnote{For a discussion and references, see Section 3.2 of the companion paper.}

So for these D-brane systems, Strominger and Vafa calculated the entropy by counting all the possible field theory states that give rise to the same values of the charges as the corresponding black hole solution had. This number agreed with the Bekenstein-Hawking entropy formula for the black hole: suggesting that here indeed was a microscopic derivation of the formula.

\begin{figure}
\begin{center}
\includegraphics[height=9.5cm]{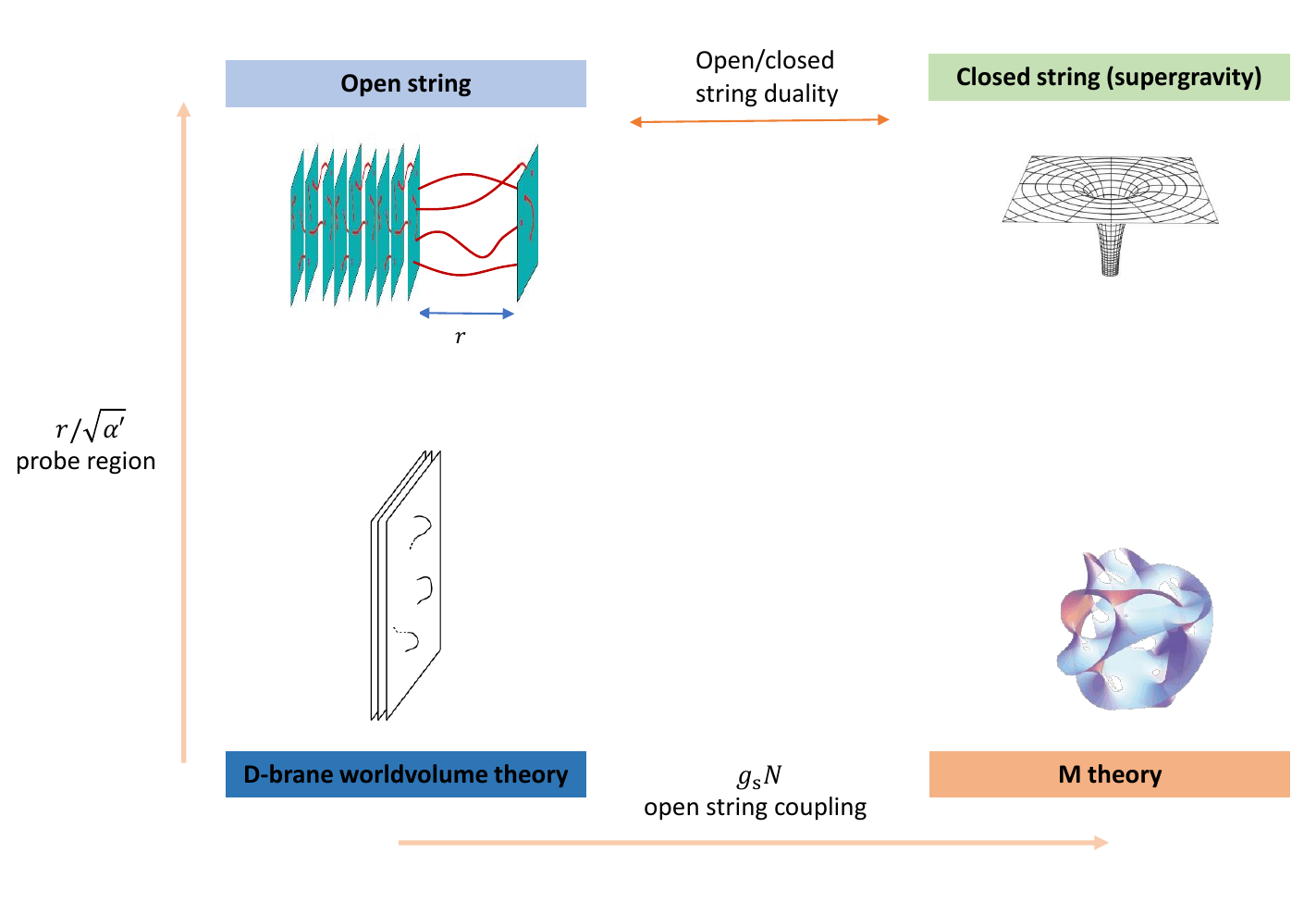}
\caption{\small Pictorial representation of the Strominger-Vafa argument: the open string coupling increases to the right and probe distances increase going up.  The top-right corner represents a massive semiclassical black hole, observed from large distances; at the top left, we find a system with heavy D-branes, probed by another D-brane at a large distance; between these D-branes, strings are attached that quantum mechanically interact. At the bottom left, there are fluctuating D-branes that are stacked closely together and described using quantum field theory. The bottom-right corner represents a still unknown non-perturbative quantum gravitational system. \emph{Source images}: Wikimedia.}
\label{SVdiagr}
\end{center}
\end{figure}

The argument is summarised in Figure \ref{SVdiagr}. It involves three related theories: supergravity in the top-right corner, as the low-energy limit of closed string theory or M-theory; open string theory in the top-left corner; and the quantum field theory on the D-branes in the bottom-left corner. The bottom-right corner, labelled `M theory', served to motivate Strominger and Vafa's work but played no explicit role in their argument. The M theory itself is in fact at this point still only a conjectured and not fully articulated non-perturbative formulation of closed string theory; we will return to this issue in our Conclusion (for more on M-theory, see also Section \ref{de}). In the Figure, the open string coupling is given by the product $g_{\sm s} N$ of the string coupling constant $g_{\sm s}$ and $N$, the number of D-branes: $g_{\sm s} N$ is often referred to as the `'t Hooft coupling'.

The motivation behind the relation between the black hole and D-brane systems is the correspondence between open and closed strings that is known as `open-closed string duality'.\footnote{Open-closed duality in perturbative string theory originated in work by Hikaru Kawai, David Lewellen, and Henry Tye (1986); see Section 2.2 of the companion paper.}
The conjectured duality  relates closed string BPS states with given charges to open string BPS states stretched between D-branes. The D-branes are the sources of the open string charges. This is the duality between the top-right and top-left corner of Figure \ref{SVdiagr}, which Joseph Polchinski had made plausible through a one-loop calculation of the tension of the D-branes. The idea was that the forces exerted between two parallel D-branes, understood to be due to open strings stretched between them, could also be considered as due to the exchange of a closed string, i.e.~a gravitational interaction.

At small separation, the open string states between the D-brane system and a D-brane probe are also described by the worldvolume theory on the D-branes (that is, a quantum field theory living on the worldvolume of the D-branes). 
In particular, the BPS open string states with given values of the charges are the ground states of the worldvolume theory for a number of D-branes that source those charges. Furthermore, since each D-brane has one unit of charge, the {\it number} of D-branes, and their kinds, together determine the total charge and thus the state. 
The probe distance should not affect the number of states counted for the system, thus the entropy calculation could be done in the regime of small separation between D-branes (the bottom-left corner in Figure \ref{SVdiagr}).

By using these identifications between string charges, D-branes and black holes, Strominger and Vafa were able to reduce the problem of black hole entropy counting to the problem of calculating the degeneracy of the ground state of a quantum field theory associated with D-branes. This counting could be performed due to earlier work by Vafa on the D-branes' worldvolume quantum field theory (see e.g.~Vafa, 1996), which is a two-dimensional quantum field theory with conformal invariance (a `CFT'). The spacetime of this CFT was the intersection spacetime of the D1 and the D5-branes, with one space and one time dimension.\footnote{For details, see `Conceptual Analysis of Black Hole Entropy
in String Theory', Section 3.2.2.} The degeneracy of the states with fixed energy in a generic, two-dimensional CFT, was known to be determined by the Cardy formula, which gives the degeneracy of states as a function of a quantity called the `central charge', $c$, and the energy level, $n$. In the case at hand, Strominger and Vafa calculated the central charge and energy level to be:
\begin{equation}\label{centralC}
c=6\left(\frac{1}{2}Q_{\tn F}^2+1\right),~~~~n=Q_{\tn H}~,
\end{equation}
here written in terms of the D-brane electric charges $Q_{\tn H}$ and $Q_{\tn F}$ that correspond to the charges of the appropriately related black hole. The Cardy formula then gives the following expression for the entropy associated with the degeneracy of the D-brane system:
\bea\label{fieldTR}
S_{\tn{D-brane}}=2\pi\,\sqrt{\frac{nc}{6}}=2\pi\,\sqrt{Q_{\tn H}\left( \frac{1}{2} Q_{\tn F}^2+1\right)}~.
\eea
Since one only expects the macroscopic black hole solution to be valid for large values of the charges $Q_{\tn H}$ and $Q_{\tn F}$, one simply drops the factor of 1. The result was the same as the classically computed Bekenstein-Hawking entropy-area formula:
\bea\label{HEF}
S_{\tn{BH}}=\frac{\mbox{Area}}{4G_{\tn N}}=2\pi\,\sqrt{\frac{Q_{\tn H}Q_{\tn F}^2}{2}}~,
\eea 
in units where $k_B = \hbar  = c = 1$, and in the second equation we have also set $G_{\tn N}=1$. Thus, in the appropriate regime of large charges, Strominger and Vafa found a precise match between the Boltzmann entropy of the D-brane configuration and the horizon entropy of the black hole solution.

\section{The Ontology of Black Hole Microstates}\label{ocdual}

In the previous Section's review of the Strominger-Vafa calculation, we described it as featuring ``two distinct systems: a classical black hole in supergravity [...] and a configuration of D-branes." These distinct systems, nevertheless,  
exhibited matching values for the relevant quantity: the entropy. Yet the relation that pairs the systems (depicted at the top of our diagram in Figure \ref{SVdiagr}) is motivated by the putative open/closed string {\em duality}. It was this putative duality that Polchinski (1995) used when he first considered D-branes as the corresponding  weak coupling manifestations  of $p$-branes. This raises the question: If the two systems are related by a duality (albeit at different values of the coupling), should they then be considered as, after all, the `same' system?

 This question has three aspects that we will discuss here. \\
  \indent (a): Must any two systems related by a duality be the same? We will answer `No, not always.' \\
 \indent (b): Did string theorists believe, or do they now believe, that the two systems in the Strominger-Vafa argument are the same? We will answer `Broadly, yes.'\\
 \indent (c): Is there a rigorous duality between open and closed strings in the Strominger-Vafa scenario? We will answer `Apparently not': and this will lead in to the next Section, on emergence.
 
 As to (a), the answer obviously depends on the meaning of `duality'. Our own view is that: (i) a duality is an isomorphism between two physical theories; but (ii) since isomorphism is a formal notion, two dual theories can have different subject-matters, and so be about distinct, though suitably isomorphic, systems.\footnote{This account of dualities is developed by De Haro (2017b);  De Haro and Butterfield (2018). See also Dieks, Dongen and De Haro (2015), De Haro (2017a) and Castellani and De Haro (2018), particularly on the relation between duality and emergence.} The recent philosophical literature has discussed under what further conditions the two systems can be judged identical, further seeking to identify under what conditions the dual theories are to be considered  {\it physically equivalent}.\footnote{See De Haro (2017b), Fraser (2018), Huggett (2017), Butterfield (2018), Read and M\o ller-Nielsen (2018).}

This topic obviously bears on whether the Strominger-Vafa argument counts as an explanation of black hole entropy. For if the two systems can be thought of as being the same, then there is little reason to doubt that the argument is explanatory. But it is by no means obvious that they  are in fact the same---and if not, one may indeed wonder whether \emph{black hole} entropy has been explained by the D-brane state-counting. String-theory critic Lee Smolin, upon learning of the state counting by Strominger and Vafa through a seminar on the subject by Juan Maldacena in 1996, put the question of physical equivalence as follows: ``The thermodynamical properties of the two systems are identical. Thus, by studying the thermodynamics of extremal branes wrapped around the extra dimensions, we can reproduce the thermodynamic properties of extremal black holes." Still, one may ask whether ``a genuine explanation of the entropy and temperature of black holes" had been provided, since ``the piles of branes are not black holes, because the gravitational force has been turned off." Smolin dubbed the view that the two models are physically inequivalent, the
 ``pessimistic point of view", which holds that ``the relationship between the two systems is probably an accidental result of the fact that both have a lot of extra symmetry."
The opposite view, that the models are physically equivalent, was dubbed the ``optimist" viewpoint, namely the view that black holes ``can be understood" as D-branes and that the symmetries ``simply allow us to calculate more precisely."\footnote{Smolin (2006:~pp.~139-141). Gravity physicist Ted Jacobson voiced similar reservations: ``[B]efore AdS/CFT, there was this wonderful state-counting [...] but the connection for me with black holes was really indirect, because they were weak coupling phenomena, and there wasn't actually a black hole there, it's just a stack of D-branes. If you turn the coupling constant up [...] non-renormalisation theorems from supersymmetry tell you that the state counting result must be preserved" and thus one could ``make a black hole"; yet even if the numbers for the entropy matched, there was no certainty that a quantum description of black hole (from which one could for instance infer that information was not lost in black hole evaporation) had resulted, Jacobson suggested  (interview with S.~De Haro, Seven Pines Symposium, Stillwater MN, 20 May 2017). Similar criticism was raised by quantum gravity scholar Carlo Rovelli, who compared the string result to a related result in loop quantum gravity: ``Black holes thermodynamics is definitely a success of string theory, and in my
opinion, the strongest evidence for its physical relevance. A similar success can be
claimed by loop gravity. Both successes are partial in my opinion. The string derivation
is still confined to, or around, extreme situations, as far as I know, and since it
is based on mapping the physical black-hole solution into a different solution, it fails
to give us a direct-hand concrete understanding of the relevant black hole degrees of
freedom, as far as I can see. The loop derivation of black hole entropy gives a clear
and compelling physical picture of the relevant degrees of freedom contributing to
the entropy, but it is based on tuning a free parameter to get the correct Bekenstein-Hawking entropy coefficients, and this does not sound satisfactory to me either'' (Rovelli 2013: p. 16). } To sum up: the assessment of ontology bears on whether an explanation for the black hole entropy can be claimed.

In assessing the physical equivalence of the two models, we will first address how string theorists have expressed themselves on this issue. It would immediately be clear why string theorists believe that the Strominger-Vafa argument is explanatory, if a standard string-theoretic interpretation shows that the two systems are reasonably considered to be the same system, at different values of the coupling. So, we first need to establish whether string theorists have in the past believed, or today believe, that the two systems in the Strominger-Vafa account are in fact the same. We can then ask to what degree that belief is justified, and how the degree of justification may affect an assessment of explanation.

\begin{figure}
\begin{center}
\includegraphics[height=7cm]{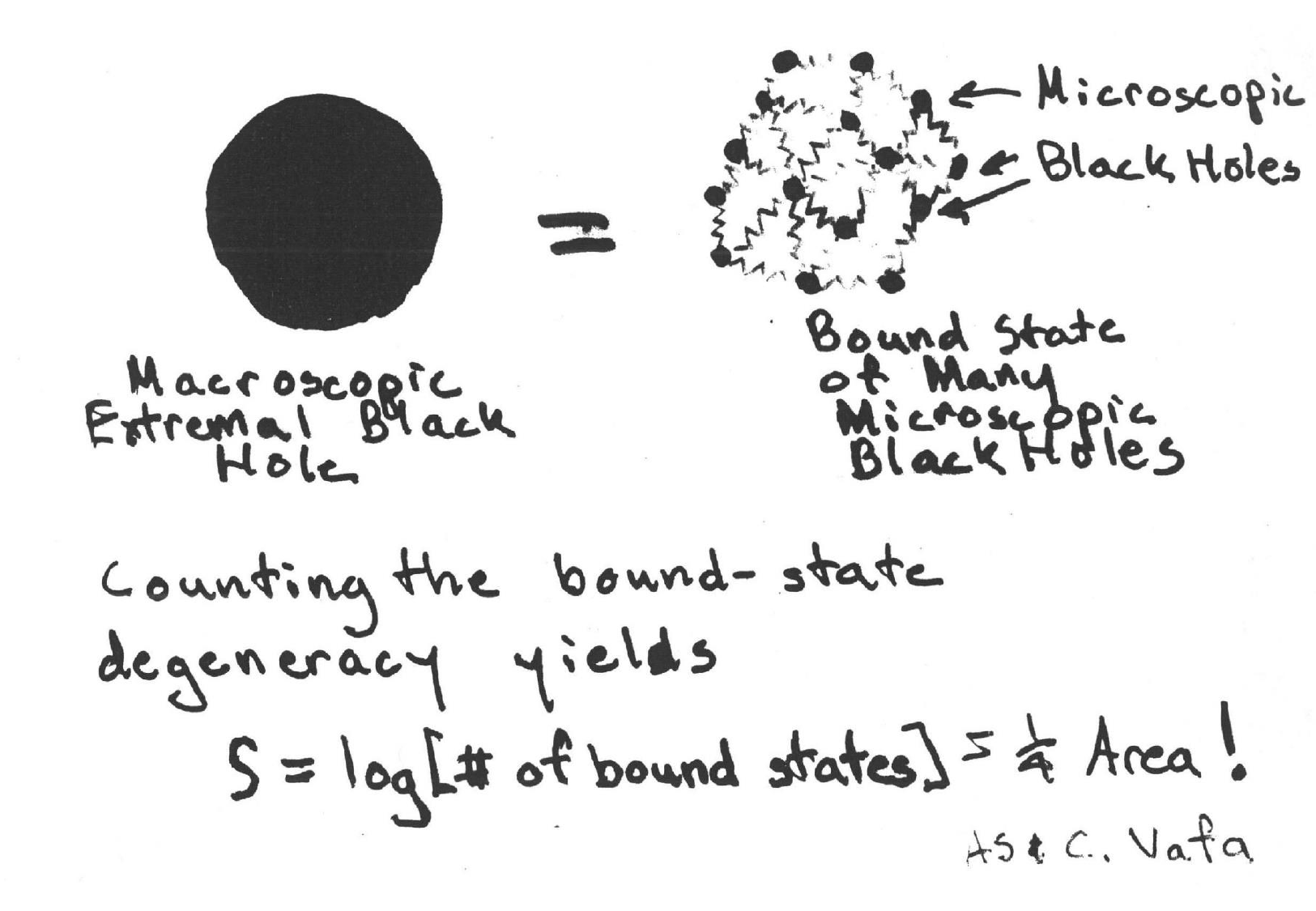}
\caption{\small Excerpt of transparency of a lecture by Andrew Strominger on ``String theory and the Bekenstein-Hawking black hole entropy" (Harvard University, October 1997), which captures the Strominger-Vafa analysis. The equality sign between the ``macroscopic" black hole and the microscopic system leaves little doubt that Strominger believed these were ontologically the `same' (with ``microscopic black holes" meaning D-branes). \emph{Source}: A.~Strominger.}
\label{stroscan}
\end{center}
\end{figure}

So we turn to question (b). Did string theorists believe that the two models are physically equivalent, i.e.~that the two systems are the same?
In their 1996 article, Strominger and Vafa spoke of ``objects" that in one regime for the coupling are captured by the ``correct physical picture'' of a ``large semiclassical black hole''---while in the other regime, ``the physical picture [...] as a supersymmetric K3 cycle [i.e.~the world-volume field theory account of the bottom-left corner in Figure \ref{SVdiagr}] is the correct [...] description." They wrote that they are dealing with ``a BPS state'' that is ``described'' in one regime of parameter values as a bound state of wrapping D-branes, which ``transforms into a hole in spacetime" in the other regime.\footnote{Strominger and Vafa (1996:~p.~103).} Even if these statements lack the precision of philosophical parlance, they do suggest that the authors took themselves to be dealing with \emph{one} object---``a BPS state"---that has two descriptions, each valid in its own range of parameters (see also Figure \ref{stroscan}). A contemporary and authoritative review of the subject is even more explicit:  ``the extremal $p$-brane in supergravity and the D$p$-brane are two different descriptions of the same object'', and these two descriptions are ``complementary."\footnote{Aharony et al.~(2000:~p.~199).}
Indeed, when asked in 2018, Vafa confirmed that he regarded the black hole and the D-brane configuration as ``the same object at different values of the parameters, at weak coupling and at strong coupling''\footnote{Interview with S.~De Haro, Harvard University, 30 November 2018.}---and Strominger expressed a similar view: ``yes, one thing, there is one real world, and it has two descriptions.''\footnote{Strominger explains his statement thus: ``It is as if you wanted to expand the function $1/(1-x)$: if you want to understand it in the unit circle you probably use $1+x+x^2+...$. On the other hand, if you want to look at large values of $x$, [...] you expand in $1/x$ where $x$ equals infinity. Both correspond to pictures that have corrections, but they are both exactly the same, and typically one picture is good for some questions and the other picture is good for other questions.
[...] [T]he string coupling constant allowed us to go from one [picture] to the other. Often [...] if you have such a parameter, you can calculate some things with one and some things with the other, while there is no overlap. But [...] BPS protection came in to save the day, because the number can't change as a function of this parameter'' (20 November 2018, Harvard University, interview by Jeroen van Dongen and Sebastian De Haro). \label{stromingernote}}

Clearly, if this one object has two descriptions that can be reliably related, and if in particular the quantity in which one is interested is invariant as one goes from one description to the other, then the derivation of that quantity---the entropy---using field theory will convincingly give the value of the entropy of a black hole. For it is the entropy of the same object.   But is the presumption of physical equivalence entirely justified, when we consider the actual relations between the two theories in the argument? To answer this question, we first have to look at whether there is a full duality that makes the argument work. So, we must discuss the role of the conjectured open-closed duality. How does this influence the physical equivalence of the right- and left-hand side of our diagram in Figure \ref{SVdiagr}?

Thus we turn to question (c). To the best of our knowledge, the present consensus is that there is {\em not} a rigorous duality between open and closed strings, in the case that Strominger and Vafa considered. Although there are cases in string theory where open and closed strings {\it are} believed to be dual to each other, with supporting evidence for the claim---two examples are AdS/CFT and topological string theory\footnote{Maldacena (1998b); Gopakumar and Vafa  (1999). One should keep in mind that the status of AdS/CFT is actually still that of a \emph{conjectured} duality, even if much evidence to support it has meanwhile been amassed; for a recent review, see Ammon and Erdmenger (2015).}---this is not known to be the case in the Strominger and Vafa scenario. Here, there is only an isomorphism of the perturbative expansion of the open and closed string theories, and current expectations are that the two do not match at the non-perturbative level.\footnote{On open-closed duality, see e.g.~Khoury and Verlinde (2000). Recent generalisations of AdS$_3$/CFT$_2$ that add an (irrelevant) deformation seem to offer prospects for `going beyond the near-horizon approximation' of AdS/CFT. The relation between open and closed strings may be then a duality after all, even in the black hole asymptotic region (Smirnov and Zamolodchikov, 2017; McGough, Mezei and Verlinde, 2018; Guica, 2018). Also, in recent work, Haco et al.~(2018) have hinted at the existence of conformal field theories for Kerr black holes. However, these results are still far from conclusive. If open and closed strings are dual to each other also in the asymptotic region of the black hole, some of our conclusions should be modified. Yet, such modifications would be straightforward (and we will indicate them at several places), since the situation would then be as in AdS/CFT, with which we will (in Section \ref{adscft}) compare the Strominger-Vafa scenario in its current interpretation.\label{ocduali}}
A full isomorphism, hence a duality, is only expected when 
``there is a complete decoupling of the supergravity description from that of the gauge theory", in the words of Clifford Johnson's textbook on D-branes.\footnote{Johnson (2003:~p.~243).}

String theorists realised in late 1997 that this decoupling happens when one focusses on the near-horizon limit in the black hole spacetime. In that case, one can ignore additional open or closed strings that generally can be expected to propagate around the black hole or D-branes: these would break the supersymmetry that is needed to secure the duality between the right- and left-hand side of Figure \ref{SVdiagr}. Indeed, when taking this near-horizon limit, open-closed duality reduces to a special case of Maldacena's celebrated AdS/CFT correspondence.\footnote{See e.g.~Aharony et al.~(2000).}
Yet, without taking a near-horizon limit---in other words, in the more general situation as in the Strominger-Vafa analysis---there is no certainty of a secure exact duality. There is only an approximation of a duality. So any  general inference from a duality (subject no doubt to some extra conditions, so as to `go beyond formal isomorphism') to physical equivalence does not apply.\footnote{Subsequent and more recent developments confirm the above picture. For example, the `OSV conjecture' (Ooguri, Strominger, and Vafa, 2004) relates the elliptic genus of the four-dimensional black hole to the partition function for a `topological subsector' of string theory (on the OSV conjecture, see also Section 4.2.4 of the companion paper.) This and similar proposals (e.g.~Vafa, 1998; Katz, Klemm and Vafa, 1999; Dijkgraaf, Vafa, and Verlinde, 2006; Haghighat et al.~, 2016\label{DVV}) are related to but do not instantiate the kind of open-closed string duality that would allow the identification between the full black hole and the corresponding configuration of D-branes, as intended in question (c) above. The OSV conjecture does not count as a duality between the full black hole and the D-brane system: for it \emph{only} maps the elliptic genus, and not the {\it full} physical partition function. It thus only captures BPS states (which do allow one to calculate the entropy of the black hole) but not other aspects such as the form of the black hole metric.}
Thus we conclude that there is no reliable evidence that the Strominger-Vafa set-up involves a duality between open and closed strings, other than approximately.

Note, however, that this weakening of the ontological relation between left and right need not undermine the validity of the identification of entropies. First of all, the same number for the entropy (the same {\it function} of the black hole charges) has been found on both sides: in one case, it was obtained by using a microphysical account of entropy; in the other, by the geometrical calculation of the area in the Bekenstein-Hawking entropy formula.\footnote{The form of this function, Eq.~\eq{HEF}, is sufficiently non-trivial that neither its functional form nor its numerical coefficients are dictated by symmetries or dimensional arguments. This is even more so for other black hole entropy functions that were calculated shortly after Strominger-Vafa. See Section 4.2 of the companion paper.} Furthermore, the BPS nature of the D-brane ground states and the
extremal nature of the $p$-brane solution  
do ensure that the entropy counts should not change when going from the theory on the left to the theory on the right by changing the coupling.\footnote{Horowitz, Strominger and Maldacena (1996:~p.~151) wrote about the Strominger-Vafa microstate counting: ``[D]ue to the special character of the BPS states involved, there is a sense in which these calculations `had to work'.''} It is then the numerical equality and  functional agreement secured by the invariance due to the fact that both systems have a mass that is equal to the charge---which follows from supersymmetry---that yields the explanatory power of the Strominger-Vafa analysis.\footnote{Our reasoning here is strictly concerned with the Strominger-Vafa argument as originally formulated. If one analysed more recent literature, one would be likely to discover a strengthening of the argument. For example, recent proposals such as the OSV conjecture and its cognates (see footnote \ref{DVV}) provide microscopic expressions for the entropy that are valid beyond the leading semi-classical approximation, and for whole classes of black holes---not just for a single example (see also Section 4.2.4. of the companion paper).} Yet, is supersymmetry alone sufficient for us to trust the identification of the entropies? As we saw, critics, at least, seem to disagree. Arguably, one should prefer to have a more direct relation
between the systems at hand. What could such a relation look like?

There are other inter-theoretic relations, other than physical equivalence, imaginable between the models on the bottom left- and top right-hand sides of Figure \ref{SVdiagr}, and supported by an approximate open-closed duality.  Johnson discusses a Reissner-Nordstr{\"o}m-type black \emph{p}-brane, just like the Strominger-Vafa case, and points out that its `Ramond-Ramond' (RR) charge should be considered as due to D-branes, since these ``actually are the \emph{basic sources} of the RR fields''; the \emph{p}-brane is to be considered ``as `made of D-branes' in the sense that it is actually the field due to $N$ [D-branes], all located at $r=0$." 
The left- and right-hand sides of  Figure \ref{SVdiagr} are still, in Johnson's words, ``two complementary descriptions'',\footnote{Johnson (2003:~pp.~241-242).} while his larger perspective suggests 
that the D-branes, extrapolated to strong coupling, should be thought of as \emph{parts, or constituents, of} the $p$-brane. He also points out another important feature that our discussion so far has set aside. Namely: while both theories in the top row of Figure \ref{SVdiagr} are only valid at weak string coupling $g_{\sm s}$, the theory on the right further requires that the number of D-branes $N$ \emph{is large}, so that  the `t Hooft coupling $g_{\sm s}N$ is large. The notion of a large $N$ limit, where $N$ is some measure of the number of degrees of freedom, reminds one of the thermodynamic limit of statistical mechanics and of the emergence of classical behaviour in large quantum systems. 

So this  suggests that, even though Figure \ref{SVdiagr}'s left and right theories 
are not known to be related by the isomorphism of an exact duality, they may perhaps be related by an \emph{emergence relation}. That is: can one think of the relation between the D-brane system and the black hole as an example of emergence? We pursue this question in the next Section.

\section{Emergence of Black Holes in String Theory}\label{SVemergence}

In this Section, we will discuss whether the notion of emergence can be applied to the Strominger-Vafa scenario. We will first, in Section \ref{concEm}, offer a straightforward conception of emergence. Sections \ref{emSV} and \ref{emwhat} then contain our central discussion of emergence in the Strominger-Vafa scenario. Finally, Section \ref{adscft} discusses emergence in AdS/CFT and compares this to the Strominger-Vafa scenario.

\subsection{The conception of emergence}\label{concEm}

In the philosophical literature,\footnote{See for example Bedau (1997:~p.~375), Bedau and Humphreys (2008:~p.~1), Butterfield (2011:~section~1.1.1), Humphreys (2016:~p.~26).} there is a widespread view of emergence as a `delicate balance' between {\it dependence}, or rootedness, and {\it independence}, or autonomy. Roughly speaking, dependence means that there is a {\it linkage} between two levels or theories: usually, a macroscopic and a microscopic theory. Independence can be taken to mean novelty in the macroscopic theory with respect to the microscopic theory. In addition, the literature sees the {\it physically significant} cases of emergent behaviour as those in which the linkage relation between the levels is {\it robust}. Robustness means that the novelty at the top level (the emergent behaviour at that level) is relatively independent of, or insensitive to, the details of the bottom level. For the purposes of this paper, we will take the condition that determines the linkage between the microscopic and the macroscopic theory to be the limit of a large number of degrees of freedom of the system, while keeping appropriate quantities fixed. Thus, it is indeed highly reminiscent of a thermodynamic limit.

String theory authors have sometimes used the term `emergence' rather liberally, and its occurrence in the technical literature on duality can be confusing.\footnote{See for example Dom\`{e}nech et al.~(2010), in which it is argued that gauge fields are `emergent' in the AdS/CFT duality. For a nuanced treatment of emergence in the context of quantum gravity, see Linnemann and Visser (2018).}   
At other times, an emergence relation might be inferred, but is not made explicit.  
For example,  Gary Horowitz, Juan Maldacena and Andrew Strominger, in a publication shortly after the Strominger-Vafa paper,  identified D-branes and black hole systems in a way that was highly suggestive, but ultimately left open whether a concrete notion of emergence applied to the black hole system. They spoke of a one-to-one correspondence since the supergravity black hole ``may be uniquely decomposed into a collection of D-branes, anti-D-branes [i.e.~oppositely charged D-branes] and strings, whose numbers we denote $(N_1,N_{\bar 1},N_5,N_{\bar 5},n_R,n_L)$ [...]. These numbers are defined [...] by matching thermodynamic properties of the black hole (under variation of the asymptotic parameters) to the thermodynamic properties of a collection of $(N_1,N_{\bar 1},N_5,N_{\bar 5},n_R,n_L)$ non-interacting branes, anti-branes and strings." The black holes ``can be viewed as `built up' of branes, anti-branes and strings"---the latter were ``fundamental".\footnote{Horowitz, Maldacena and Strominger~(1996:~pp.~152 and 155).} Still, it is not clear if one must imagine the black hole description to `emerge'  from the quantum microscopics.

\subsection{Emergence in the Strominger-Vafa scenario}\label{emSV}

In the Strominger-Vafa analysis, we are dealing with the comparison of a microphysical state-counting with a thermodynamic entropy. 
So, can one indeed identify the semiclassical black hole as emergent?

\begin{figure}
\begin{center}
\includegraphics[height=5.5cm]{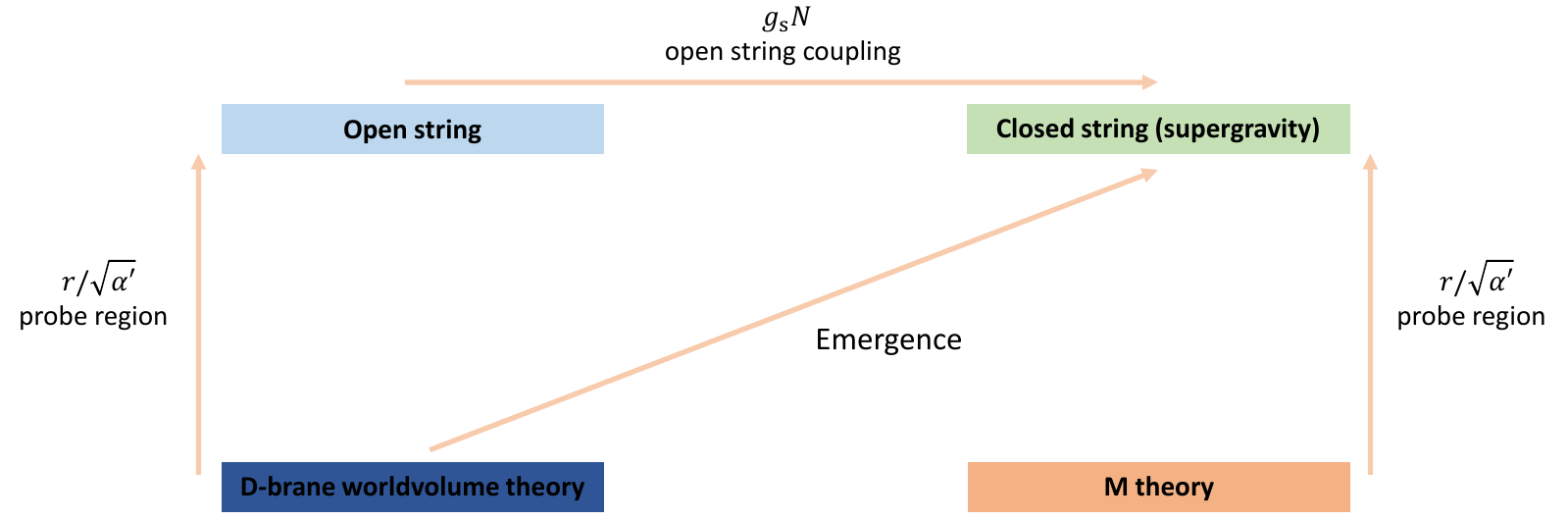}
\caption{\small Emergence of the black hole from the worldvolume D-brane system in the limit of large number of D-branes $N \gg 1$ (or large central charge) and large probe distance $r/\sqrt{\alpha'} \gg 1$. }
\label{Adiabatic}
\end{center}
\end{figure}

As was pointed out in Section \ref{SVarg},  
the states of the Strominger-Vafa black hole are determined by their charges, $Q_{\tn H}$ and $Q_{\tn F}$. These are large in the regime in which supergravity is valid:
$Q_{\tn H}\gg Q_{\tn F} \gg 1$.\footnote{See Eq. (3.6) in our companion paper `Conceptual Analysis of Black Hole Entropy in String Theory'.}
Since each D-brane carries an elementary unit of charge, this is the limit of a high number of degrees of freedom in the D-brane regime: it corresponds to taking the number of D1- and D5-branes to be large, i.e.~$N_1, N_5 \gg 1$, as one moves from the top left to the top  right in Figure \ref{SVdiagr}. Furthermore, the black hole system is an effective description, considered at large distances: it is a low energy approximation to a presumed M-theory system in bottom-right, where high energy modes have been discarded
as one moves to large distances and the top-right corner of Figure \ref{SVdiagr}. Both vectors combined suggest that emergence may occur between the top-right corner of Figure \ref{SVdiagr}, relative to the bottom-left corner, where the entropy calculation is performed: as indicated in Figure \ref{Adiabatic}.

As we stated earlier, emergence typically links a macroscopic theory with a microscopic theory. The Strominger-Vafa argument relates the black hole solution to the D-brane solution at small separation and weak coupling---in which a microstate calculation is done. 
So interpreting the Strominger-Vafa argument as exhibiting emergence in this sense entails that the (semi-)classical gravity description is interpreted as `emergent' and `macroscopic'. We will elaborate the point in what follows.

The linkage relation between the world-volume theory and supergravity  
involves taking the central charge in the world-volume theory to be large. 
In a conformal field theory, the central charge, $c$, is an extensive measure of the number of degrees of freedom
of the system\footnote{See for example Di Francesco, Mathieu and Senechal ~(1997:~pp.~135-136). A free boson field has central charge 1, and a free fermion field has central charge 1/2. The central charge is determined by the two-point function of the stress-energy tensor. The stress-energy tensors of uncoupled systems simply add, so that the central charge of the total system is the sum of the central charges of the subsystems. So, the central charge of $N$ free bosons is $N$, and that of
$N$ free fermions is $N/2$.}---hence the resemblance with a thermodynamic limit. For the Strominger-Vafa system, the central charge and energy level are given by Eq.~\eq{centralC}. 
Thus taking $c$ and $n$ large indeed corresponds to taking $Q_{\tn F}\gg1$, $Q_{\tn H}\gg1$.

Strominger and Vafa explicitly used the limit of large $c$ in their comparison between the D-brane world-volume theory and the supergravity theory. 
On the one hand, the entropy associated with the degeneracy of the D-brane system is computed from the Cardy formula, in Eq.~\eq{fieldTR}, that is   derived in the  D-brane or weak coupling regime  $g_{\sm s}N  \ll 1$ (which translates into  $Q_{\tn F}^2 \ll Q_{\tn H}$ or $c \ll n$).\footnote{This follows from the proportionality of the string  coupling to the ratio of the charges, $g_{\tn s} \sim Q_{\tn F} / Q_{\tn H}$, and from setting $N = Q_{\tn F}$; see Eqs. (3.5) to (3.7) in our companion paper.}
On the other hand, in the supergravity or strong coupling   regime  $g_{\sm s}N \gg 1$ and $g_{\sm s} \ll 1$---these conditions together  are equivalent to    $Q_{\tn F}^2 \gg Q_{\tn H} \gg Q_{\tn F} \gg 1$ or $c \gg n \gg 1$---one can compute the horizon area, which is then identified with the thermodynamic Bekenstein-Hawking entropy, Eq.~\eq{HEF}.
Strominger and Vafa assumed that the Cardy formula is also valid in the supergravity regime---when the central charge is large, i.e.~$c\gg 1$---and they found that in this regime 
the right-hand sides of Eq.~\eq{fieldTR} and Eq.~\eq{HEF}, respectively the Boltzmann entropy and the Bekenstein-Hawking entropy, are equal.\footnote{See Section 3.1 in our companion paper for a detailed discussion of the different regimes. For supersymmetric BPS states, the Cardy formula can be applied in the supergravity regime  because of the invariance of the entropy under changes of the coupling (see Section 2). For non-supersymmetric 2d CFTs the range of validity of the Cardy formula can still be extended to the supergravity regime, if the theory has a large central charge and a small number of low-energy states (see Hartman, Keller and Stoica, 2014).}
Strominger and Vafa finally said that in the presence of many D-branes, or, in other words, ``for sufficiently large charge", ``the \emph{correct physical picture} of the objects we discuss is as a large semiclassical black hole." The picture provided by string perturbation theory should ``break down" in the regime of large charges,\footnote{Strominger and Vafa (1996:~p.~103), emphasis added.} 
 even if the entropy count is still valid due to its invariant nature.

Hence, the formal analysis of the number of degrees of freedom as a function of the charges is certainly suggestive of emergence: the black hole is novel and macroscopic and may thus be expected to emerge along the direction of the tentatively-inserted upward diagonal arrow in Figure \ref{Adiabatic}.\footnote{In a 2018 interview, Cumrun Vafa stated that as you turn up the coupling, the D-branes turn into a black hole, but ``there is no particular point where it happens"; to say that the black hole ``emerges" would be ``the correct way to state it. It emerges in the sense that it becomes more and more similar to what we call a black hole. [...] Particles and black holes are not that different, except [...] in this case a large amount of charge'' (C. Vafa, interview with S. De Haro, Harvard University, 30 November 2018).} So far as the entropy count is concerned, the emergence seems robust, and so physically significant, because the macroscopic result is independent of the details of the microscopic theory (such as the precise location of the open strings on the D-branes, the type of D-brane configuration, etc.).\footnote{The entropy is the same for  any  combination of D1-, D3-, and D5-branes with momentum,  intersecting along the $S^1$, as long as they preserve the right amount of supersymmetry and the total Ramond-Ramond charge is $Q_{\tn F}$ (see Section 3.2.2 in our companion paper).}

\subsection{Emergence of what? 
}\label{emwhat}

But notwithstanding Section \ref{emSV}'s discussion: in order to claim emergence, in the sense of the philosophical literature outlined in Section \ref{concEm}, one needs to be more precise about what it is that emerges. For, strictly speaking, the Strominger-Vafa argument shows only that the black hole entropy  
matches the number of degrees of freedom in a corresponding microscopic theory, for large $N$---if you wish, that the entropy `emerges'. But this is different from showing that a physical system emerges. The argument matches a single quantity between the two theories, but does not show that the entire black hole system emerges out of the D-branes. In particular, it does not show how the black hole {\it metric} emerges from the D-brane world-volume theory. One could argue that for us to speak of `full' emergence, or even to say the Strominger-Vafa argument `explains' \emph{black hole} entropy, this further step must also be achieved.

So does the black hole metric emerge from the D-brane system, for instance to leading order in the approximation in which this metric is valid? And how necessary is such emergence, for us to be sure that the calculated entropy corresponds to the entropy of the black hole? Unless some direct linkage between the microscopic system and the macroscopic system can be established, the calculation might not be an explanation of  entropy as a thermodynamic property of the macroscopic system. For as we pointed out in Section \ref{ocdual} (question (a)), one might be comparing properties of systems that ontologically have nothing to do with each other. Yet, we also learned (Section \ref{ocdual}, question (b)) that some physicists are ``optimists'' about the relation between the systems: for example, Strominger and Vafa themselves. And we observed (Section \ref{ocdual}, question (c)) that although we have no solid ground to think that the relation between open and closed strings is a duality, there is room for a close ontological relation---quite possibly of a relation of emergence.

Furthermore, the consistency of the string theory programme's assumptions would seem to require that the black hole metric emerges, at least to leading order in the low-energy approximation used. After all, we know that D-branes give a good description of interacting open strings, and the latter are related---at least in perturbation theory---to closed strings by the relation between open and closed strings, which is close to being a duality. Finally, supergravity is the low energy limit of closed string theory. So if these inter-theoretic relations are correct, there should be at least a perturbative match between the black hole metric and the geometry of the effective moduli space of the D-brane system, i.e.~the geometry of the space in which the D-brane world-volume fields take values, which, under the duality, are reinterpreted as spacetime coordinates.

Indeed, in August of 1996 Michael Douglas, Daniel Kabat, Philippe Pouliot and Stephen Shenker outlined a general approach that does invoke a notion of emergence of the metric that is best explicated as we did above, i.e.~in terms of novelty (namely, a new spacetime) and linkage (in this case, the linkage is a case of `derivation'). They wrote: ``In the latter [field theory] description, space-time emerges as a derived concept, the low energy moduli space [i.e.~space in which fields take values] of a supersymmetric gauge theory.''\footnote{Douglas, Kabat, Pouliot and Shenker~(1997:~p.~87).} Here, the authors~were referring to attempts at establishing the emergence of a classical spacetime  (the top-right corner of Figure \ref{SVdiagr}) from the theory on the D-brane world-volume, in the bottom-left corner. However, the theories considered  were not specifically aimed at the Strominger-Vafa black hole. So can one identify that emergence relation also in the Strominger-Vafa scenario?

Fourteen months after the Strominger-Vafa paper, Douglas, Polchinski and Strominger (1997) wrote a preprint in which, using techniques from Douglas, Kabat, Pouliot and Shenker~(1997), they tried to match the moduli space metric of its D-brane world-volume theory to the black hole metric. More precisely: they treated the D1-brane as a probe in the background of the D5-branes and calculated the effective action that prescribed its motion. From this action they read off the D-brane moduli space metric, order by order in the loop expansion in the D-brane world-volume theory. Comparing this to the black hole metric of Strominger and Vafa, they found agreement in the case of the tree-level and one-loop terms of the moduli space metric  at large distances (i.e.~for small values of $\sqrt{\a'}/r$, exactly as one expects from Figures \ref{SVdiagr} and \ref{Adiabatic}; the tree-level and one-loop terms in the metric are then order 1 and order $\a'/r^2$, respectively). 

Alarmingly, they found disagreement at two loops, i.e.~at order $\a'{}^2/r^4$ in the metric---but Juan Maldacena, in a lecture at \emph{Strings '97} in Amsterdam, reported that the second order term would also work out.\footnote{Maldacena (1998a); on the recovery of the full black hole metric, see e.g.~also Dijkgraaf, Verlinde and Verlinde (1997b); Chepelev and Tseytlin (1998).} Among the systems he studied was a D1-brane probe intersecting a stack of D5-branes, as in the Strominger-Vafa case (he also considered a number of near-extremal systems). The effective action for the motion of the D1-brane was obtained upon integrating out the open string degrees of freedom. Since this effective action for the motion of the brane was then reinterpreted as prescribing a supergravity background and metric, classical geometry was to follow from D-brane quantum field theory after integrating out the high energy degrees of freedom and focussing on the resulting effective action. These efforts clearly strengthen the identification of the Strominger-Vafa scenario as an example of emergence, as illustrated in the arrow of Figure \ref{Adiabatic}.

Ultimately, the natural goal of such reconstruction efforts would have been to recover the {\it entire} black hole metric from the D-brane world volume theory, with possible quantum corrections due to fluctuations. At the same time, they led Maldacena to intuit a ``correspondence'' that needed ``to be explored" further: in his lecture in Amsterdam,  he was on the threshold of his AdS/CFT duality, which followed from studying the near-horizon limit of the D1-D5 system.\footnote{See Maldacena  (1998a:~p.~26); the AdS/CFT result is found in Maldacena (1998b); for a brief description of its history, see our companion paper `Conceptual analysis of black hole entropy in string theory', Section 5.} Soon, attention poured into that subject, and the counting of black hole states, reconstructions of classical spacetimes and related problems were predominantly studied in that context. In the case of the AdS/CFT correspondence,  the open and closed string descriptions are fully dual, so one expects a full retrieval of the geometry from the quantum system, as was indeed later shown in various examples and approximations.\footnote{See e.g.~De Haro, Skenderis and Solodukhin (2001).}

In the next Section, we will discuss the ontology and emergence of the black hole in the context of AdS/CFT. But for now, what can we conclude about the Strominger-Vafa scenario? We submit that the discussion so far establishes that emergence {\it is} part of the relationship between the scenario's D-brane and black hole systems. The successes in identifying terms of the metric, along with the entropy, suggest that one is quite likely justified in ontologically relating
the D-brane system and black hole, even if they are not the same systems in the philosophical or technical sense of the word `same'--- as we saw in Section \ref{ocdual}, the open-closed string duality is, so far as we know, only approximate: so that we lack grounds for {\it identifying} the two systems.
However, it is reasonable to regard the black hole as emergent from the D-branes, as we have argued here. Can we also be more precise in identifying the kind of emergence relation involved? 
\\
\\
Philosophers often distinguish between ontological and epistemic emergence. The intended contrast is, roughly, between  emergence `in the world', and emergence merely in our description of the world. So the question arises which of these is illustrated by the Strominger-Vafa scenario. To answer this, we adopt an explication of ontological vs. epistemic emergence introduced by one of us (De Haro 2019b, especially Section 2.2). Recall Section \ref{concEm}'s account of emergence as a linkage between a `bottom' microscopic theory (or `level') and a `top' macroscopic one, that shows the top theory to be rooted in (dependent on) the bottom theory, while also exhibiting novel behaviour or properties (usually in a robust manner, i.e.~largely insensitive to details at the bottom level). One can add to this construal of emergence the idea of {\em interpretation maps}. Such a map is a function mapping bits of a theory (paradigmatically: words, and other syntactic items) into the theory's domain of application, i.e.~appropriate objects, properties and relations in the physical world.\footnote{Beware of the differing jargons from mathematics and from physics: the range, in mathematical jargon, of the interpretation map is a subset of the theory's domain of application (physics jargon), i.e.~part of the physical world.} Ontological emergence is a matter of the interpretation maps for the bottom and top theories not matching in their ranges: in particular, the top theory's domain of application, i.e.~the range of its interpretation map, has one or more items (objects or properties) that are {\em not} in the bottom theory's domain---ontological novelty indeed.\footnote{More precisely, we are here (and below) concerned with the emergence of the black hole, in terms of its spacetime properties and its dynamics. Our usage thus agrees with Wong's (2010:~p.~7) definition: ``Ontological emergence is the thesis that when aggregates of microphysical properties attain a requisite level of complexity, they generate and (perhaps) sustain emergent natural properties.'' However, use of the phrase `ontological emergence' is unfortunately not uniform in the literature (witness the various uses considered in O'Connor and Wong (2015:~Section 3.2): viz.~ontological emergence as `supervenience', as `non-synchronic and causal', as `fusion', to name a few). Our account of ontological emergence is metaphysically pluralist, in that while it recognises that emergence is a matter of novelty in the world, it does not point to a {\it single} metaphysical relation (e.g.~supervenience, causal influence or fusion) as constitutive of ontological emergence.} On the other hand, in epistemic emergence the domains \emph{do} match. 

We can put this in mathematical jargon. We think of the linkage as a map $\ell$ from the bottom theory to the top theory. Then $\ell$ being non-injective expresses the idea  that  the novel behaviour or properties in the top theory are insensitive to details at the bottom. (Think of coarse-graining; or more generally, of collective variables.) If we also write $i_b, i_t$ for the interpretation maps, then epistemic emergence requires that $i_t \circ \ell = i_b$, i.e.~there is a commuting triangular diagram of functions. On the other hand, in ontological emergence the diagram is non-commuting: ran$(i_t) \neq$ ran($i_b$).

Given this distinction between ontological and epistemic emergence, our question about the Strominger-Vafa scenario is: Is the emergence of the black hole from the D-brane system ontological? In other words: do novel properties `in the world' arise when following  Figure \ref{Adiabatic}'s diagonal emergence arrow from the D-brane to the supergravity description? Although a conclusive answer would require full details about interpretation maps etc., we propose that the answer is `Yes'. 
Recall the fact we stressed above: that the duality between open and closed strings is not known to be exact---at least, outside of the AdS/CFT correspondence (cf. the next Section). An inexact duality would give good prospects of novelty, as one moves rightwards as $g_{\sm s}N$ increases. Indeed, the black hole description has the striking novelty that it is a gravitational description, while the D-brane theory has no gravity.\footnote{As we mentioned before, evidence points to the {\it absence} of a duality in the asymptotic region far away from the black hole, as we have assumed in this Section: but the evidence is inconclusive, with some recent hints that the Strominger-Vafa scenario could be extended to be a duality, and so  the question remains technically open. If it turns out that there is a duality between open and closed strings that is also valid away from the near-horizon limit, then this conclusion
would of course change, but then the emergence discussed in the next paragraph is still valid (see also Section \ref{adscft}).}
Thus, the black hole as a gravitational system emerges, as one turns up the open string coupling $g_{\sm s}N$.

Furthermore, there may be ontological emergence vertically on the left also, in the D-brane system---the D-brane world-volume theory is an effective description of the open string theory that is obtained by integrating out massive degrees of freedom---and vertically on the right, as the probe distance is altered.\footnote{This is not a `mere' integrating out of degrees of freedom, as when e.g.~one defines centre-of-mass coordinates for a body, which is a mere change of variables, and for which one does not speak of emergence. Rather, one integrates out in such a way that a description of a novel system or situation 
becomes possible.} 
Compare this, for example, with the similar, simpler and more familiar case of the emergence of classical thermodynamics from statistical mechanics by coarse-graining: which can be held to be ontological. Then, in case of an exact duality at the fundamental level, moving diagonally in Figure \ref{Adiabatic} would still yield an emergent scenario, by virtue of its vertical component.

\subsection{Emergence of black holes in AdS/CFT}\label{adscft}

In this subsection, we address how to think of emergence for a black hole 
in the context of AdS/CFT. Here, we assume, {\em ex hypothesi}, an exact duality between the closed string, gravitational side (the bulk AdS side) and the boundary CFT side, which is the quantum field theory on the D1-D5 brane intersection, in the limit in which the open and the closed strings are decoupled.
Emergence in AdS/CFT was analysed in detail in earlier work;\footnote{See Dieks, Dongen, De Haro~(2015), De Haro (2017a), and Castellani and De Haro~(2018).} so here we will first summarise that discussion. Then we will briefly compare it with the  Strominger-Vafa scenario discussed above, and end with some remarks about information loss.\\
\\
Two of us, in Dieks et al.~(2015), argued that, in cases of an exact duality, there can be no ontological emergence in the sense of one side of the duality emerging from the other (for AdS/CFT, the putative emergence usually considered is that the gravity side emerges from the CFT). This idea of emergence `across a duality' will be the first of  three cases we will consider: call it `case (i)'.

Our argument was 
that, since two dual theories are isomorphic formulations of a single theory, there is no room for ontological emergence, which requires an asymmetric inter-theoretic relation between two theories (including all their physical quantities). For  isomorphism precludes such an asymmetric relation: two theories cannot be isomorphic in all their states, quantities, and dynamics if there is a non-injective linkage relation between them (the map $\ell$: cf. Section \ref{emwhat})---as is required for emergence. In other words, there are two formal requirements in tension: the isomorphism and the non-injectivity relations between the sets of states, quantities, and dynamics.
Indeed, Andrew Strominger similarly would not regard one side of the AdS/CFT duality as more `fundamental' than the other, and is ``reluctant" to imagine gravity to emerge in that case: ``that would be giving one of the pictures a more fundamental status---but expansion in $x$ and expansion in $1/x$ [of the function $1/(1-x)$ around $x \rightarrow 0$ and $x\rightarrow \infty$] are both equal, yet to say that one is emergent from the other [...] gives one of them a more fundamental status.''\footnote{Andrew Strominger, interview with Jeroen van Dongen and Sebastian De Haro, Harvard University, 20 November 2018; see also note \ref{stromingernote}.}

In the language of De Haro (2019b), emergence  requires the existence of a `bottom' theory and a `top' theory, that are linked by a surjective and non-injective linkage map. For example, Ernest Nagel (1961:~p.~357) discusses the reduction of thermodynamics to statistical mechanics, whereby temperature is connected  
to the mean kinetic energy of gas molecules.
But when two such theories are isomorphic, the non-injectivity requirement is violated, because an isomorphism is of course injective. This is why the existence of a duality map precludes this type of emergence. Two theories that are isomorphic are much more closely connected than an emergence relation allows.

What does this argument against emergence `across a duality'  mean in the case of AdS/CFT? Taking this duality to be  formulated (as usual) as pairing fields in semiclassical supergravity theory on the gravity side, with conformal field theory correlation functions: the argument means that neither side is emergent from the other when moving along the horizontal direction. In particular, gravity is not emergent from the field theory. This argument would not apply to the Strominger-Vafa case, however, given the absence of an exact duality there.\\
\\
So much by way of discussing case (i): whether  one side of a duality can emerge from the other side. But there is another dimension to this discussion: the dimension of length or energy, as against coupling-strength (or open vs. closed)---associated, as we have already seen, with the vertical, as against horizontal, directions of our Figures. 

The first point to make here is that on the basis of the AdS/CFT duality, one may infer the existence of a more fundamental, microscopic theory on the gravity side (viz.~a quantum theory of closed strings, perhaps formulated via M-theory), given the existence of a microscopic theory on the boundary. Then, should the isomorphism apply in the black hole case: there would indeed also be a quantum black hole object at strong `t Hooft coupling (bottom right of our Figures). 

But the more general point is of course that in the context of a duality, emergence might still occur `individually'---vertically---within each side of the duality (cf. Dieks et al., 2015; De Haro, 2017). This is the second of three cases we will consider: call it `case (ii)'. 

For example, we might envisage that the supergravity theory (with a black hole solution) emerges out of the underlying closed string theory on the gravity side. If so, we are viewing supergravity (with possible quantum corrections, which are small sufficiently far away from the black hole singularity) as an effective, low-energy theory. Of course, this is a classical theory emerging out of a quantum one, and so
it reminds one of the familiar classical-quantum correspondence relation: we will say more about this  in the next Section.

There remains work to be done here in order to explicate the precise details of emergence: in terms of our notation, an explicit linkage relation $\ell$ needs to be worked out. 
But let us assume that that work can be done. (Here, the expectation would be that the black hole is seen as a coherent, macroscopic state of closed strings and other objects in closed string theory.) Then since in AdS/CFT there is an exact duality, this behaviour can be mapped, using the duality map, to a similar emergent behaviour in the CFT (such behaviour is usually associated with the existence of a new non-perturbative vacuum with novel properties in the CFT). Thus, it is possible to get emergence `on each of the two sides of the duality', with the emergent behaviour being mapped one into the other. 
To sum up case (ii): the two sides of a duality could each be a case of emergence, with the duality mapping the two cases exactly to each other.\\
\\
Another possible case of emergence---call it `case (iii)'---is that we combine emergence on just one side---one of the emergence arrows from the previous paragraph, i.e. on the left-hand side---with the duality map: so that we regard supergravity as emerging from the (non-perturbatively defined) CFT.\footnote{For completeness, we should add that there are two additional cases that (i)-(iii) do not cover. First, our emphasis has been on ontological emergence: for some details about {\it epistemic emergence}, see Castellani and De Haro (2018). Second, two dual theories might be given different interpretations, implying  ontological `novelty' of each with respect to the other. But this is not a case of emergence:
rather, it is a mere change of interpretation, which ``does not introduce an appropriate relation between the [emergent] system and its comparison class'' (De Haro, 2017:~p.~118). For, as we said in Section \ref{emwhat}, emergence requires a linkage map that is non-injective, and so cannot be bijective.}
In that case, one need not in fact assume the existence of a fundamental gravity theory; gravity itself would be an emergent phenomenon. We will return to this point in our conclusion, Section \ref{conclsn}. 

This case (iii) is reminiscent, of course, of  Figure \ref{Adiabatic} in Section \ref{emSV} (see also the end of Section \ref{emwhat}), with the option to leave out the bottom-right corner.\footnote{It is further reminiscent of proposals by Erik Verlinde (2011); see also their discussion in Dieks, Dongen and De Haro (2015: Section 4).}  But it is important to remember that in the case of the Strominger-Vafa black hole, one does \emph{not} have an asymptotic AdS space,  and probably \emph{not} a duality.  
If, however, it turns out that there {\it is,} after all, a duality between open and closed strings, 
then the above scenarios (ii) or (iii) could indeed be regarded as candidates for emergence scenarios in which the black hole is emergent within the context of an exact duality. \\
\\
Given the above options, what might we infer about information loss? If the entire black hole were described by an exact duality, then, as we stated, one could infer the existence of a microscopic gravity theory. Since the dynamics of the field theory that links the states is known to be unitary, that aspect would carry over from the field theory to the gravity theory under the duality, implying that the microscopic black hole object has unitary evolution laws as well---and suggesting that there would be no information loss. Similarly, in the absence of a microscopic gravity theory, i.e.~in case gravity is a fully emergent phenomenon as in case (iii), one can still infer that microscopic evolution is unitary, as the CFT dynamics is unitary. 

However, in the situation of the Strominger-Vafa black hole, i.e.~in today's absence of a duality for the full black hole solution, such an inference is far from certain.
In the words of Andrew Strominger, when there is a duality, 
``in principle we can calculate anything we want using expansion in $x$ or expansion in $1/x$; [...] [they] are on equal footing.'' 
However, ``in the presence of the black hole, you have to be careful  [...]. 
The bulk gravity theory has an asymptotically flat region [i.e.~is not asymptotically AdS], and the field theory of the whole system, including the asymptotically flat region, is something that we actually used to not understand at all.'' According to Strominger, some progress has been made since 2016 towards an exact, full formulation of open-closed duality that would also potentially capture the black hole.\footnote{See e.g.~Smirnov and Zamolodchikov (2017); McGough, Mezei and Verlinde (2018); Guica (2018).} But 
since such results are far from fully secured, it is premature to pass judgment on information loss. Thus, to quote Strominger again:  
``suppose we have just the gravity description and we know all about string theory and we are really good at doing world sheet diagrams, can we imagine that we could, even when there is a black hole, calculate everything using the field theory? That is Hawking's problem, who claimed that you can not: without resolving the information paradox, we do not know whether there is a way to do that calculation or not." Indeed, a full duality, or even better, a microscopic evaporation scenario, and ``not just the state-counting'', would be necessary to really answer the question of information loss.\footnote{Andrew Strominger, interview with Jeroen van Dongen and Sebastian De Haro, Harvard University, 20 November 2018.} 

\section{The Correspondence Principle}\label{corrp}

We have seen that inter-theoretic relations play an essential role in the Strominger-Vafa counting of black hole microstates. In Sections \ref{SVarg} and \ref{ocdual}, we discussed the role of the conjectured open-closed string {\it duality} in the counting. Then, in Section \ref{SVemergence}, we reviewed how the Strominger-Vafa argument illustrates the inter-theoretic relation of {\it emergence}. 

In this Section, we will look at how inter-theoretic relations of {\it correspondence} enter the Strominger-Vafa argument: and we will relate this to Bohr's correspondence principle, which played a key role in the development of quantum theory. 
First, in Section \ref{Bcorr}, we briefly review Bohr's correspondence principle and its historical context. Following Robert Rynasiewicz, we distinguish two versions of the principle. Then, in Sections \ref{freqth} and \ref{de} respectively, we will draw analogies between their historical roles and   the development of string theory in the context of the Strominger-Vafa calculation. 

\subsection{Bohr's correspondence principle}\label{Bcorr}

The string theory programme aspires to create a quantum theory of gravity, with limited input: on the one hand, classical gravity needs to be retrieved in an appropriate limit, and on the other, essential quantum features are expected to be in the final theory. In this sense, string theory's programmatic goal reminds us of the challenges physicists faced in the early twentieth century, when classical electrodynamical and mechanical theory on the one hand, and empirical quantum results on the other, had to be combined in a consistent theoretical framework. Of course, in today's search for quantum gravity, the quantum input is largely rooted in theoretical rather than empirical knowledge. But one may still find revealing resemblances between these attempts at theory construction---regardless of one's view of the larger epistemic status of string theory. 

One of the essential problems that the early quantum theory sought to address was the structure of the hydrogen atom and its spectral lines. We will compare the role of this problem to the role that extremal and near-extremal black holes  played in attempts in the 1990s at quantum gravity. Both are problems where the `borders' of key theories (for black holes: field theory, thermodynamics and relativity) are probed; and for both, a number of constraining and guiding relations were available, such as the Balmer or Hawking spectra. 
In Niels Bohr's efforts at theory construction, the \emph{correspondence principle} and Paul Ehrenfest's \emph{adiabatic theorem} played important roles.\footnote{For a concise history of the old quantum theory, see Jammer (1989); more recent historiography is found in e.g. Aaserud and Kragh (2015); Dongen et al.~(2009); Katzir et al.~(2013);~Navarro et al.~(2017).} Both of these can also be recognised in string theorists' efforts on black holes.

When speaking of `the' correspondence principle in the old quantum theory, one is immediately confronted with the question: \emph{what} correspondence principle exactly? Bohr was a notoriously verbose and associative author. In particular, pinning down his precise version of the correspondence principle has proven elusive.\footnote{Paul Ehrenfest used his contribution to the 1921 Solvay conference to try to do just that; see Ehrenfest (1923); Bohr's own immediately preceding contribution is Bohr (1923).} Nevertheless, for the purposes of our discussion we will follow the insightful treatment of Robert Rynasiewicz, who has traced different versions and uses of the principle.\footnote{Rynasiewicz (2015); for an alternative treatment, see e.g.~Bokulich (2008). We stress that the typical modern textbook construal of the correspondence principle, namely that one guesses the quantum Hamiltonian $\hat H(\hat Q,\hat P)$, as being the classical function $H$ of $\hat Q$ and $\hat P$, is {\em not} the notion of correspondence in play in this discussion. Compare the next footnote. An insightful general discussion of the correspondence principle in relation to its historical role in the old quantum theory is Radder (1991).} Rynasiewicz has identified two implicit uses of the correspondence principle in the old quantum theory: its role as a `frequency theorem', and the much more heuristic `downward extrapolation project'. 

The \emph{frequency theorem} states that the classical orbital frequency of the valence electron coincides with the frequency of the emitted radiation in the limit of orbits with large quantum numbers. It does so numerically, even if conceptually the situation remains unclear. That is: in this limit the electron can be treated as a charged oscillating source while in the quantum picture (even at large $n$) the radiation is due to a transition, not an oscillation.

The frequency theorem also prompted the correspondence principle to be identified with a heuristic strategy that Rynasiewicz labelled \emph{the downward extrapolation project}. This strategy sought to use classical equations to offer justifications and relations applying to typical quantum phenomena, in regimes in which classical physics was expected to be invalid.  Precise or even consistent formulations of this strategy were elusive; and its operation seemed to only bear fruit in the hands of Bohr and his associates. 
It began by observing that the classical orbital motion of the electron can be described by a Fourier decomposition of its trajectory: 
\begin{equation}
x (t) = \sum_{\tau} X_{\tau} \cos{2 \pi (\tau \omega_{\tau} t )}. \label{fourier}
\end{equation} 
Here, $x$ is the electron's position, $\tau$ a running integer and $\omega_{\tau}$ the electron frequency of Fourier mode $\tau$.
The Fourier coefficients $X_{\tau}$ were taken as the key to a host of quantum phenomena:  they were used as a measure for the intensity of spectral lines, and to determine selection rules and, when suitably generalised to three dimensions, polarizations. For instance, when  $X_{\tau}=0$ in the classical expression (\ref{fourier}), this was interpreted as an indication of the absence of a particular spectral line and transition between two quantum states $n$ and $n'$ with $\tau=n'-n$. For us, the main point is that this type of use was \emph{not} restricted to regimes of high quantum numbers $n, n'$, but extrapolated to low quantum numbers and the full quantum regime: `downward extrapolation'. Werner Heisenberg, for example,  depended on this strategy when formulating matrix mechanics: he focussed on the set of $X_\t$ as the main observable quantities when inserting
Eq.~(\ref{fourier}) into the Bohr-Sommerfeld quantization rule, thus deriving the quantum commutation relation for matrix operators in 1925.\footnote{In subsequent elaborations of the quantum theory (as in e.g.~Dirac 1930), \emph{the} correspondence principle gradually became identified with the `frequency theorem', and no longer with a strategy of extrapolating to the quantum regime on the basis of classical relations; see Rynasiewicz (2015).} 

In the next two Subsections, we discuss whether these two versions of the correspondence principle---the frequency theorem, and downward extrapolation---have analogues in the Strominger-Vafa scenario and the larger string effort to construct a quantum theory of gravity.   

\subsection{The frequency theorem for black holes}\label{freqth}

The Strominger-Vafa argument contains a transition between the classical and the quantum realm that is analogous to the `frequency theorem' of Bohr. For here, too, a full fundamental quantum system is taken to return a classical system in the limit of large quantum numbers. 

In the Bohr case, this limit involves going to large quantum number and large distances in the quantum atom, i.e.~$r / r_{e} \gg 1$, with $r_{e}$ the Bohr radius. Analogously, in the full open string theory, considered at large distances compared to the string length, one takes the limit of large $N$ (or strong `t Hooft coupling $g_{\sm s}N$) to arrive at the supergravity black hole. The latter, too, is probed at large distances $r / r_{\tn S} \gg 1$, with $r_{\tn S}$ the Schwarzschild radius. In both the Bohr case and open string theory, the quantum system is taken as in principle valid at all energy scales: it is the `fundamental' system.

\begin{figure}
\begin{center}
\includegraphics[height=6.5cm]{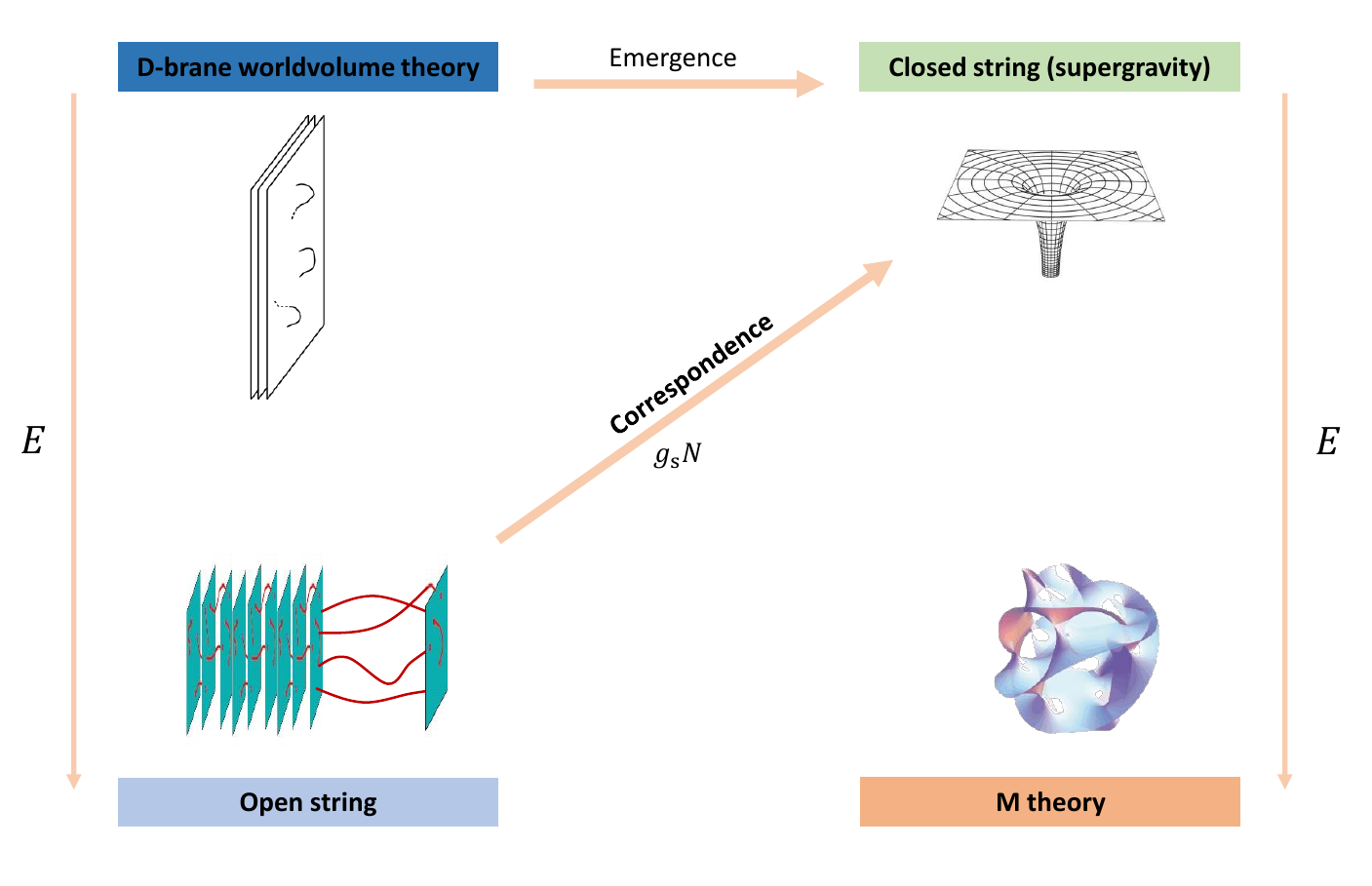}
\caption{\small The correspondence arrow, which represents a limit in which $g_{\tn s}N \rightarrow \infty$ and the systems are related by open-closed string duality.  Energy scales are flipped when going from the bottom left to the top right as an aspect of the open-closed string duality; the energy scale is depicted along the vertical. The emergence arrow is also depicted in this diagram. The two calculations that make up the Strominger-Vafa argument itself are performed only in those corners (top-left and top-right) where $g_{\tn s}$ is small, since otherwise the theories would be poorly controlled. 
 }\label{CORRESP}
\end{center}
\end{figure}

The situation is shown in Figure \ref{CORRESP}, with the more fundamental quantum systems at bottom. (Cf. Figure 3 of Section 3.3 of our companion paper.)  The correspondence arrow relates the (quantum) open string theory with D-branes for $g_{\sm s}N \rightarrow \infty$ to the $p$-brane black hole solution of supergravity. That arrow captures that one moves to both low energies and large quantum numbers when going from the full quantum system to its classical limit (in contrast to Figure \ref{Adiabatic}, the new Figure \ref{CORRESP} has the energy scales along its vertical axis; emergence involves going to both large $N$ and large distances and is here a rightward arrow).  
On the right hand side, the regime of large $N$ is identified with the system at low energies, as high energy modes are integrated out when moving vertically up. As one moves along the correspondence arrow, the energy scales are inverted. This is also a manifestation of the open-closed string duality.

To do their calculation of the entropy, Strominger and Vafa needed one more step. They had to go to small distances, and to the D-brane world-volume theory (the top left in Figure \ref{CORRESP}). 
There, they could count states. And the result remained valid because of invariances, both in linking the D-brane system to the supergravity solution as one varies the coupling, and in linking the D-brane systems at large and small separation. These invariances are the consequence of the BPS or extremal nature of the systems, and needed to be navigated by making `adiabatic' changes. That is: variations in the coupling or probe distance would not affect the state-counting in the D-brane system, since the number was an invariant.

This logic reminds us of the methodology of Paul Ehrenfest's adiabatic principle: Ehrenfest sought to find out which variables should be quantised by finding out which remained invariant under slow, adiabatic changes to a system.\footnote{See e.g.~Ehrenfest (1916) [1968].} Clearly, both in the old quantum theory, and in the current effort to find a theory of quantum gravity, the strategy of seeking variables that remain invariant as one varies parameters of scale, coupling, energy 
or number of states, is central, particularly in the context of correspondence. In the case of Ehrenfest, the goal was to identify the variables that were to be quantised; while in the case of Strominger and Vafa, one knew which quantity should remain invariant (viz.~entropy) and the goal was to identify the appropriate theory and system for which the logic would work.

Finally, it is worth stressing that the `correspondence arrow' in Figure \ref{CORRESP} relating quantum to classical also represents Polchinski's original hypothesis relating D-branes and $p$-branes.
Indeed, that hypothesis is akin to the correspondence principle. 
For first, sending $N \rightarrow \infty$ gives many D-branes: high charges between the branes and thus highly excited strings. Then, one considers these at large distances compared to the string scale. In other words: one is looking at a quantum system in the limit of high quantum numbers, and at distances large compared to the typical quantum scale of the system, $l_s$ (compare with the Bohr radius in the quantum atom). Polchinski then used open-closed string duality to transfer to closed string theory,  
whereby high energies are mapped to low energies and vice-versa, so that they are effectively inverted one with respect to the other
(see Section 2.2, and the end of Section 3.3, of the companion paper).
The duality maps the open strings connecting the $N$ D-branes with high energy excitations to an exchange of many closed strings (or gravitons) between the same D-branes at lower energy scales.\footnote{See Figure 1 in our companion paper.} 
The collective effect of the latter is to curve space for a probe: or in other words, to produce a $p$-brane supergravity solution. 

In the chosen limit, closed string theory is indeed usually studied as a classical theory. Polchinski further showed that quantum mechanical quantities of the D-branes in the open string theory, such as their tension and charge, match the relevant $p$-brane quantities, computed classically. So, as in the case of Bohr's `frequency theorem', in which a quantum treatment numerically matches at large quantum numbers a classically computed quantity: here, the quantum treatment of open string theory at large $N$ matches a classical treatment for a non-trivial gravitational object, i.e.~the $p$-brane solution of a supergravity theory. 

Thus, we see that in both Polchinski's D-brane-to-$p$-brane scenario and in the Strominger-Vafa argument, the quantum-classical correspondence is numerically accurate in key quantities, and is exhibited for large quantum numbers. They are not, however, extrapolations to a full quantum regime on the basis of classical theory: they are not examples of `downward extrapolation'---to which we turn now. 

\begin{figure}
\begin{center}
\includegraphics[height=6.5cm]{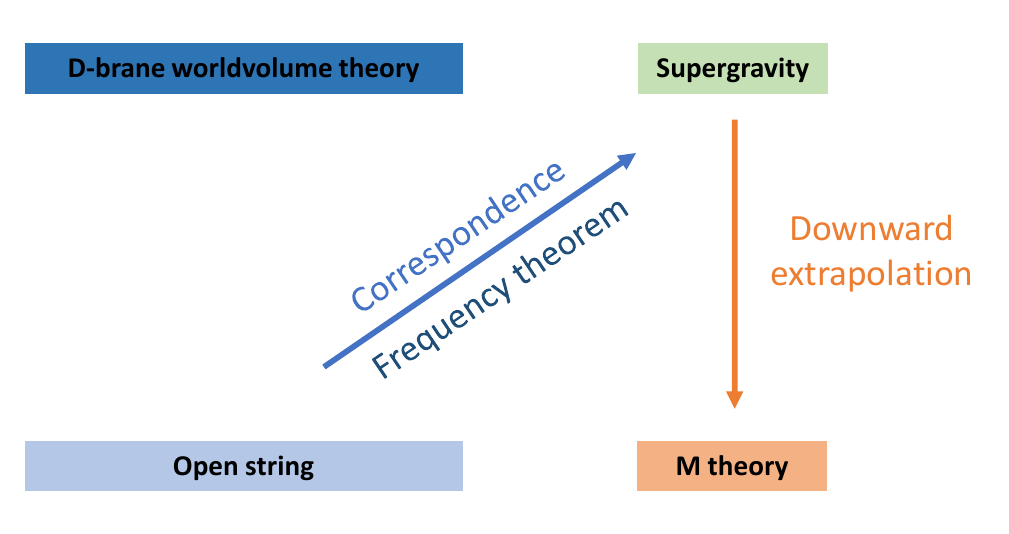} 
\caption{\small The two roles of the correspondence principle, used in the construction of the old quantum theory, illustrated in the Strominger-Vafa  and string theory context.  Frequency theorem: the quantum treatment of open string theory at large $N$ matches a classical treatment for supergravity black solutions. Downward extrapolation: inferences about the details of M-theory in low $N$-regimes, based on the behaviour of the supergravity theory in the semiclassical regime.} \label{Down}
\end{center}
\end{figure}

\subsection{Downward extrapolation and M theory}\label{de}

Do we see strategies of downward extrapolation in string theory's account of black holes? Or in other attempts at a full quantum theory of gravity in the broader string theory literature? We will first answer the latter question and then return to the concrete cases of string theory black holes and AdS/CFT.

\begin{figure}
\begin{center}
\includegraphics[height=6cm]{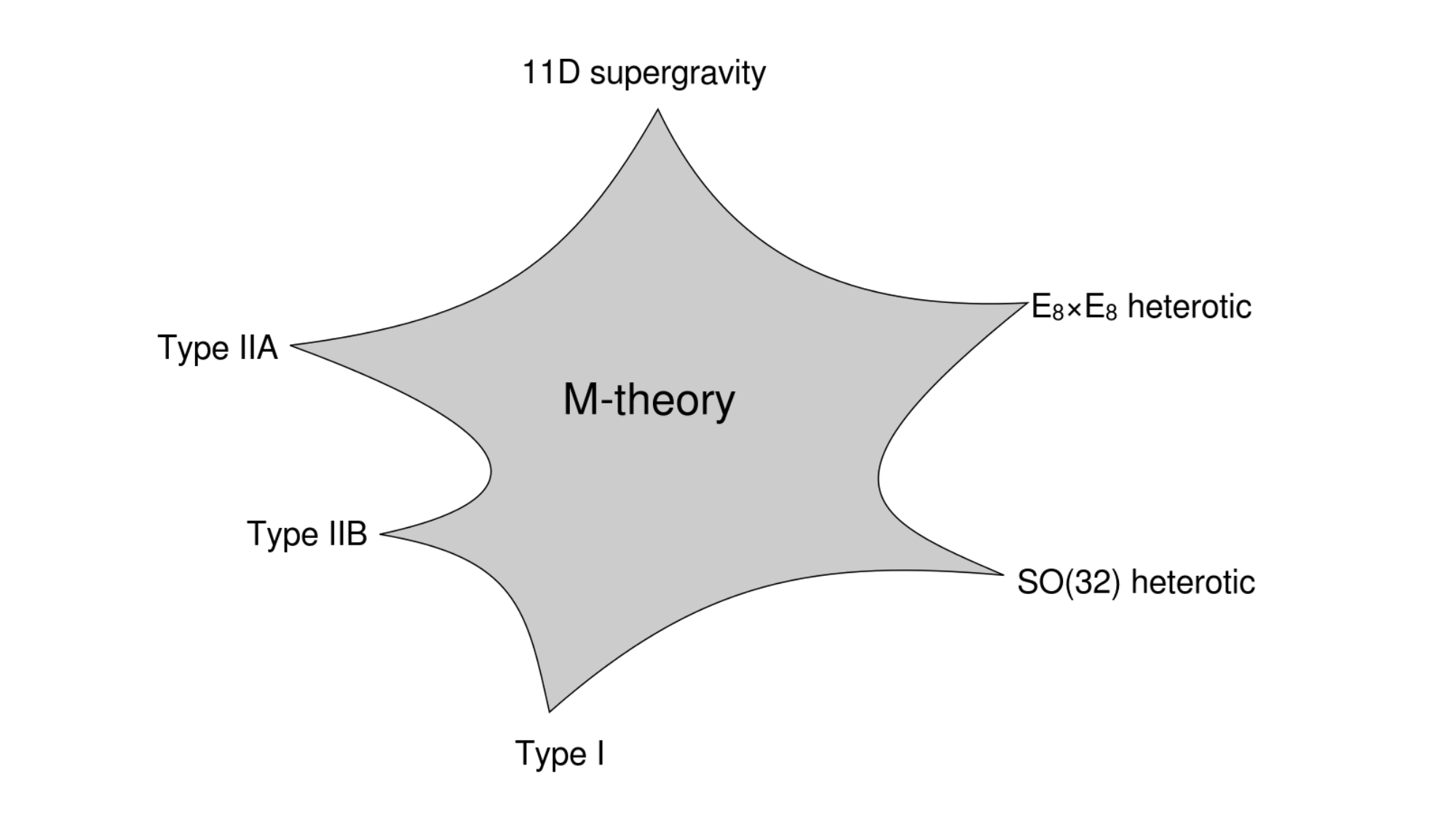}
\caption{\small M-theory, with 11-dimensional gravity in one of its corners. {\it Source}: Wikimedia. }
\label{mtheory}
\end{center} 
\end{figure}

An obvious example of `downward extrapolation' is Edward Witten's M-theory proposal itself (see Figure \ref{Down}).\footnote{The heuristic role of dualities, of guiding the search for a successor theory such as M theory, has also been explored in De Haro (2019a).} This is best illustrated by focusing on another Figure, namely Witten's familiar diagram that depicted the `web of dualities' (Figure \ref{mtheory}). At the endpoints of a web, it depicts a number of string theories that are related to one another by dualities or some other mapping relation. These theories are all considered in some simplifying limit of large quantum numbers or small value of the coupling. The conjecture is that `at the centre' of this web one will find a full quantum theory of gravity: `M-theory.'\footnote{Witten (1995).} For instance, if one were to start at the endpoint that is 11-dimensional supergravity, one could in principle go inwards, towards the full quantum mechanical M-theory regime, by increasing the energy or probing distance. Just like in the attempts of Bohr's generation, the behaviour at the full quantum level is inferred by heuristically extrapolating from the classical regime.

M-theory would be able to give a valid description of any quantum gravity process. Two months before Witten's M-theory proposal, Paul Townsend had suggested that ten-dimensional strings might descend from quantum supermembranes in eleven dimensions; in the diagram of Figure \ref{mtheory}, this would be captured by moving from the Type IIA endpoint towards the centre.\footnote{Townsend (1995). In the article, he also derived all the then known $p$-branes in type IIA string theory from the classical 11-dimensional supermembrane and five-brane.} Supergravity in eleven dimensions (the theory depicted at the endpoint at the top of Figure \ref{mtheory}) contains classical supermembranes; Townsend further introduced their quantum cousin, the `M-brane'. In an article paper entitled ``D-branes from M-branes'', he argued that certain examples of these M-branes, upon compactification on a circle, would reproduce Polchinski's D2- and D5-branes. Late in December 1995, Witten would explore some quantum aspects of the 11-dimensional five-brane, such as anomalies.\footnote{Townsend (1996); Witten (1996b). The latter also famously declared that the {\it M} of `M-theory' ``stands for magic, mystery, or membrane, according to taste'' (p.~383).} Thus M-theory was to be a theory of M2-branes and M5-branes.\footnote{Note that the labels `M2' and `M5' were only introduced later. During this period, they were usually called (as they still occasionally are) `supermembrane' and `five-brane', respectively.} 
In this way, aspects of M-theory were inferred from 11-dimensional supergravity. 

M2-branes, however, were hard to quantise.\footnote{See Wit et al. (1988, 1989). Townsend's (1995) proposal to resolve this problem was instrumental in developing the idea of M-theory itself.} Consequently, a proposal was made to build up the M2-branes out of D0-branes (that is, point particles) with non-commuting coordinates---and this proposal, in an obvious play on the historical analogy with Heisenberg's creative moment, was dubbed ``matrix theory."\footnote{Banks et al.~(1997); Dijkgraaf, Verlinde and Verlinde (1997a: p. 43).} In an appropriate large $N$-limit, the matrix coordinates would reduce again to the familiar classical spacetime coordinates. 
Thus the strategy of downward extrapolation, along with the frequency theorem, was clearly in use in string theory's effort to go from 11-dimensional supergravity and string theory towards M-theory.

Has the strategy also played a role in the string-theoretic analysis of black holes? Strominger and Vafa did not set out to formulate a full quantum gravity theory, yet one obvious example where the answer is `Yes', is the black hole information paradox.\footnote{Another example even involves an explicitly formulated ``correspondence principle for black holes and strings", discussed in late 1996 by Gary Horowitz and Joseph Polchinski (1997). They discussed a proposal by Leonard Susskind (1998) [1993] that was made before the (re)discovery of D-branes. Susskind had suggested a one-to-one correspondence between Schwarzschild black holes and string states. By changing the string coupling, he tried to match the entropy of string states with black hole entropy, as the string scale goes from below the Schwarzschild radius to above the Schwarzschild radius. Horowitz and Polchinski pointed out that D-branes should be considered as part of the system, and thus yield a corrected entropy count. Our point is that these discussions try to illuminate the full quantum gravity theory by seeking an explicit correspondence with, and heuristic expansion, on the classical. These are examples of downward extrapolation, though their exact formulations of what constitutes `correspondence' may differ from Bohr's.} 
As we saw, the Strominger-Vafa entropy counting has been taken as an argument against the inference to non-unitarity for the full quantum regime, notably by Strominger and Vafa themselves in their 1996 article. The argument is that, since a unitary theory has given a correct entropy count, a unitary account of evaporation may also be expected. As in downward extrapolation, the argument reaches beyond its premises---as Strominger also realises (see  Section \ref{adscft}).

One can also point to attempts to find the `grey body factors' of Hawking radiation, i.e.,  the factors that determine to what degree the radiating black hole is not the perfect black body emitter described by Hawking's semi-classical analysis. These factors are an example of how microscopics may deviate from semi-classical expectations. Their calculation has involved matching semi-classical absorption cross-sections using supergravity with D3 and indeed D1-D5 brane systems. So inferences about the quantum world are  rooted in numbers that are, at least partly, classically derived: as in the heuristic strategy of downward extrapolation.\footnote{Work on grey body factors is found in Maldacena and Strominger (1997a); Das and Mathur (1996).}

These developments, which closely followed the Strominger-Vafa analysis, were instrumental in the formulation of the AdS/CFT proposal.\footnote{See Aharony et al.~(2000:~pp.~200-206); our companion paper, Section 5.} 
That proposal, too, is a clear example of downward extrapolation, and of using the frequency theorem. First, a match is found between correlation functions for quantum fields in the classical gravity background of AdS-space and those of a quantum field theory on its boundary at large $N$. This match is then taken as a reason to expect that the boundary field theory can also reveal how quantum gravity 
would look non-perturbatively, away from the large $N$-limit;\footnote{See Maldacena (1998b), in particular on pp.~246-247 in its conclusion.} 
thus, its formulation resonates with the strategies of Bohr and his contemporaries

Interestingly, the two examples of correspondence are dissimilar as regards the ontologies that they connect. In Bohr's case, the classical and quantum images (or, in the jargon we introduced in Section \ref{emwhat}, `maps') of a system cannot be fully identified with one another; whereas AdS/CFT, if assumed to be an exact duality, ensures an isomorphism between the two and allows the inference to a common ontological core for the system.  For a string-theoretic black hole, the ontological relations may then be more like Bohr's case, due to the presumed approximate nature of the open-closed  string duality.

Finally, we note how in AdS/CFT emergence and correspondence are closely related, too. The emergent properties of spacetime (vertically) function as a heuristic guide in the construction of the full quantum gravity theory. As Figure \ref{Down} already illustrated in the case of the black hole, the theory that emerges from the correspondence relation is taken to suggest, by downward extrapolation, the fundamental theory of quantum gravity: in the case of AdS/CFT, duality-type inferences then offer its holographic construction as a field theory (horizontally across the duality).

\section{Conclusion}\label{conclsn}

In this concluding Section, we will not summarise the previous Sections' discussions {\em seriatim}. Instead, we will first discuss correspondence, and then return to emergence. 

In the Strominger-Vafa black hole and AdS/CFT, as in the old quantum theory, classical theory has functioned as both a constraint and a springboard. In these cases, a similar notion of correspondence has been invoked.\footnote{Another interesting comparison may be made with Erwin Schr\"{o}dinger's attempts at finding a quantum mechanical wave equation, in analogy to the optical-mechanical analogy of Hamilton; for more on Schr\"{o}dinger's attempts, see Joas and Lehner (2009). We thank Jaco de Swart for this suggestion.}
Of course, even if string theorists do not and did not have empirical input about the high-energy regimes that their theory is supposed to describe, they could orient themselves by considering history. 
Although historical analogy has hardly ever been explicitly mentioned as giving heuristic direction or justification, there were clear instances in which a proposed phrase implicitly invoked it, presumably to clarify the author's methodological strategy; examples are the cases of `matrix string theory', the `correspondence principle for black holes' of Horowitz and Polchinski (1997) and the `black hole complementarity' of Susskind et al.~(1993) and Kiem et al.~(1995). Thus, string theorists themselves implicitly but clearly played on a sense of historical continuity with  familiar earlier efforts at theory construction. Such historical resonances aid in justifying and gaining authority for their own attempts.

The Strominger-Vafa black hole, as a prominent example in string theory, illustrates one of the theory's particularly puzzling aspects: an aspect that it also shares with Bohr's uses of the correspondence principle in atomic physics. Namely, the fact that the two pictures are related by correspondence leaves it unclear whether they have the same underlying ontology. In Bohr's case, the pictures offer conflicting accounts. In the case of the string theory black hole, as we saw, the duality-like relation that connects the two regimes---field theory and supergravity---needs to be appreciated. The issue here is that its status as a full duality is, so far, insecure and undecided: which prevents a definitive judgment about ontology. Still---based on the available evidence---we have judged the systems to be 
related by an emergence relation that is a case of ontological emergence, i.e., in which the top-level theory exhibits novel properties in the world. 

So, what states have been counted in the Strominger-Vafa scenario? Are they `black hole' states?\footnote{The question itself can be seen as another example of the definitional ambiguity that the concept of a `black hole' faces; see Curiel (2019).} Our analysis suggests that, indeed, the counted states are states of a black hole, yet the properties typical of a black hole have only manifested themselves in the emergent limit.
However, answering these questions means navigating the duality relationship just outlined, and so any answer could not be final, since the status of the duality is as yet undecided. This also weakens any inferences about the issue of information loss, even though an inference to non-unitary evolution can be considered less likely given the microscopic counting of entropy in a unitary string theory.

Indeed, the Strominger-Vafa argument, by focussing on supersymmetric BPS D-brane systems and extremal black holes, built in invariances ensuring that the field theory entropy-count matches the gravity-theory's answer, as parameters vary. The match between a microscopically calculated entropy---the degeneracy of a quantum state---and a thermodynamic entropy may raise confidence that a unitary accounting of black hole evaporation will be found. The match may also raise confidence in the interpretation of the black hole horizon area as a measure for the physical object's entropy. Agreed, the calculation was performed for a highly idealised, higher-dimensional version of a black hole. Yet, additional results, such as state-countings for spinning or near-extremal black holes,\footnote{See, respectively,  Breckenridge et al.~(1997) and Callan and Maldacena (1996). The latter do not read off the entropy from the Cardy formula (cf. Eq. \eqref{centralC} et seq.), but count states explicitly, giving their derivation of black hole entropy a more statistical mechanical flavour; see our companion paper, Section 4.2.1.} and also the AdS/CFT correspondence, further strengthen such judgments.  

We have argued that a proper way to understand the relation between the gravity and D-brane systems in the Strominger-Vafa scenario is that of an emergence relation, and the supergravity black hole is itself considered to be a low energy approximation of fundamental closed string theory, or M-theory. 
However, M-theory itself never directly enters into the argument:  
the entropy counting is done in the low coupling regime of a D-brane system. 
Thus, one way to construe the Strominger-Vafa entropy calculation
is as a result that does not in fact depend on the existence of a fundamental theory of quantum \emph{gravity}. On such an interpretation, gravity only manifests itself in the emergent limit, and is not seen when the coupling is low. This may remind one of holographic scenarios that propose the gravitational force as an emergent, entropic phenomenon.\footnote{See Verlinde (2011); for a discussion of its relation with philosophical notions of emergence, see Dieks, Dongen and De Haro (2015); De Haro (2017a).}

But such vistas are beyond the immediate scope of this article. Here, we have attempted only to review the ontological and  epistemological issues associated with entropy state-counting in string theory, by studying one of its most cited exemplars. Even in the absence of empirical input about the high-energy regimes that string theory is supposed to describe, we have seen similarities  with earlier episodes in theoretical physics in which constraints and heuristics  played essential roles in the formulation of novel theory. Thus string-theoretic practice seems more familiar from the perspective of past physics than string critics who point to a lack of empiricism\footnote{E.g.~Ellis and Silk (2014).} may have realised. More generally, however: we submit that the Strominger-Vafa result raises a wealth of interpretative issues that urgently need philosophers' attention---we duly invite them to join the debate.

\section*{Acknowledgements}
\addcontentsline{toc}{section}{Acknowledgements}

We are most grateful for the generous time shared in interviews and in giving feedback to this article and its companion by Andrew Strominger and Cumrun Vafa. We further thank Erik Curiel, Peter Galison, Sean Gryb, Ted Jacobson, Jos Uffink, Erik Verlinde and David Wallace for discussions and essential assistance to our project. We are also grateful for valuable feedback from two referees. SDH's work was supported by the Tarner scholarship in Philosophy of Science and History of Ideas, held at Trinity College, Cambridge, and by a 
fellowship of the Black Hole Initiative at Harvard University; JB and JvD are very grateful for the latter's hospitality in the fall of 2018. MV is supported by the Netherlands Organisation for
Scientific Research (NWO; project number SPI 63-260).

  \section*{References}
\addcontentsline{toc}{section}{References}

\small

Aaserud, F. and Kragh, H., eds.~(2015). {\it One Hundred Years of the Bohr Atom.} Copenhagen: The Royal Danish Academy of Sciences and Letters.\\
\\
Aharony, O., Gubser, S.S., Maldacena, J.M., Ooguri, H.~and~Oz, Y.~(2000). `Large N field theories, string theory and gravity', {\it Physics Reports}, 323, pp.~183-386.  
[hep-th/9905111].\\
\\
Ammon, M. and Erdmenger, J.~(2015). \emph{Gauge/Gravity Duality. Foundations and Applications}. Cambridge: Cambridge University Press. \\
\\
Banks, T.~(2005). `Space time in string theory', pp.~311-349 in: Ashtekhar, A. (ed.), {\it 100 Years of Relativity. Spacetime Structure: Einstein and Beyond.} Singapore: World Scientific.\\
\\
Banks, T., Fischler, W., Shenker, S.H.,~and Susskind, L.~(1997). `M-theory as a matrix model: A conjecture',
  \emph{Physical Review D}, 55, pp.~5112-5128.  
  [hep-th/9610043].\\
\\
Bedau, M.A.~(1997). `Weak Emergence', {\it Philosophical Perspectives}, 11, pp.~375-399. 
\\
\\
Bedau, M.A.~and Humphreys, P.~(2008). {\it Emergence: Contemporary Readings in Philosophy and Science.} Cambridge, MA: MIT Press.\\
\\
Bekenstein, J.D.~(1972). `Black holes and the second law', \emph{Lettere al Nuovo Cimento},  {4}, pp.~737-740.
  \\
  \\
Bekenstein, J.D.~(1973). `Black holes and entropy', \emph{Physical Review D}, {7}, pp.~2333-2346.
\\  
\\
Bohr, N.~(1923). `L'application de la th\'{e}orie des quanta aux probl\`{e}mes atomiques', pp.~228-247 in: {\it Atomes et \'{e}lectrons. Rapports et discussions du conseil de physique tenu \`{a} Bruxelles du 1er au 6 avril 1921 sous les auspices de l'institut international de physique Solvay}. Paris: Gauthier-Villars.\\
\\
Bokulich, A.~(2008). {\it Reexamining the Quantum-Classical Relation. Beyond Reductionism and Pluralism}. Cambridge: Cambridge University Press.\\
\\
Breckenridge, J.C., Myers, R.C., Peet, A.W., and Vafa, C.~(1997).
  `D-branes and spinning black holes',
  {\it Physics Letters B},  391, pp.~93-98.
  [hep-th/9602065].\\
\\
Butterfield, J.~(2011). `Emergence, reduction and supervenience: a varied landscape', \emph{Foundations of Physics}, 41, pp.~920-959. 
[arXiv:1106.0704]. \\
\\
Butterfield, J.~(2018). `On dualities and equivalences between physical theories'. Forthcoming in: Huggett, N., Le Bihan, B., and  W\"uthrich, C. (eds.), {\it Philosophy Beyond Spacetime}, Oxford: Oxford University Press. [arXiv:1806.01505].\\
\\
Callan, C.G.~and Maldacena, J.M.~(1996). `D-brane approach to black hole quantum mechanics', {\it Nuclear Physics B}, 472, pp.~591-610. 
  [hep-th/9602043].\\
  \\
Castellani, E.~and De Haro, S.~(2018). `Duality, fundamentality, and emergence'. Forthcoming in: Glick, D., Darby, G., and Marmodoro, A. (eds.), {\it The Foundation of Reality: Fundamentality, Space and Time}, Oxford: Oxford University Press. [arXiv:1803.09443].\\
\\
Cecotti, S. and Vafa, C.~(1993). `On classification of N=2 supersymmetric theories', {\it Communications in Mathematical Physics}, 158, pp.~569-644. 
    [hep-th/9211097].\\
\\
Chepelev, I. and Tseytlin, A.A.~(1998). `Long distance interactions of branes: Correspondence between supergravity and super Yang-Mills descriptions',  \emph{Nuclear Physics B}, {515}, pp.~73-113. 
  [hep-th/9709087].\\
  \\
Curiel, E.~(2019). `The many definitions of a black hole', \emph{Nature Astronomy}, 3, pp.~27-34. 
[arXiv:1808.01507].\\
\\
Das, S.R. and Mathur, S.D.~(1996). `Comparing decay rates for black holes and D-branes', {\it Nuclear Physics B}, 478, pp.~561-576. [hep-th/9606185].\\
\\
De Haro, S.~(2017a). `Dualities and emergent gravity: Gauge/gravity duality', 
  {\it Studies in History and Philosophy of Science} B: {\it Studies in History and Philosophy of Modern Physics}, {59}, pp.~109-125. [arXiv:1501.06162].\\
\\  
De Haro, S.~(2017b). `Spacetime and physical equivalence'. Forthcoming in: Huggett, N. and W\"uthrich, C. (eds.), {\it Space and Time After Quantum Gravity}, Cambridge: Cambridge University Press. [arXiv:1707.06581]. \\
\\
De Haro, S.~(2019a). `The heuristic function of duality', {\it Synthese},
196 (12), pp.~5169-5203. 
[arXiv:1801.09095].\\
\\
De Haro, S.~(2019b). `Towards a theory of emergence for the physical sciences'. {\it European Journal for Philosophy of Science,} 9: 38.\\ 
\\
De Haro, S.~and Butterfield, J.~(2018). `A schema for duality, illustrated by bosonization',  pp.~305-376 in:  Kouneiher, J. (ed.), {\it Foundations of Mathematics and Physics One Century After Hilbert}, Cham: Springer. [arXiv:1707.06681].\\
\\
De Haro, S., Solodukhin, S.N.,~and Skenderis, K.~(2001). `Holographic reconstruction of space-time and renormalization in the AdS/CFT correspondence',   \emph{Communications in Mathematical Physics},  {217}, pp.~595-622.   [hep-th/0002230].\\
\\
Dieks, D., Dongen, J. van, and De Haro, S.~(2015). `Emergence in holographic scenarios for gravity', {\it Studies in History and Philosophy of Science} B: {\it Studies in History and Philosophy of Modern Physics}, 52, pp.~203-216. [arXiv:1501.04278].
\\
\\
Di Francesco,~P., Mathieu,~P.,~and Senechal,~D.~(1997). \emph{Conformal Field Theory}. New York: Springer. \\
\\
Dijkgraaf, R., Vafa, C.,~and~Verlinde, E.~(2006). `M-theory and a topological string duality'. Preprint, 15 pp. [hep-th/0602087].\\
\\  
Dijkgraaf, R., Verlinde, E.,~and~Verlinde, H.~(1997a). `Matrix string theory',
  \emph{Nuclear Physics B}, {500}, pp.~43-61. 
[hep-th/9703030].\\
\\  
Dijkgraaf, R., Verlinde, E.,~and~Verlinde, H.~(1997b).
   `5-D black holes and matrix strings',  \emph{Nuclear Physics B}, {506}, pp. 121-142. 
   [hep-th/9704018].\\
\\
Dirac, P.A.M.~(1930). {\it The Principles of Quantum Mechanics}. Oxford: Clarendon Press.\\
\\
Dom\`{e}nech, O., Montull, M., Pomarol, A.,  Salvio, A.,~and~Silva, P.J.~(2010). `Emergent gauge fields in holographic superconductors',  {\it Journal of High Energy Physics},  {1008}, 033.
[arXiv: 1005.1776].\\
\\
Dongen, J. van, Dieks, D., Uffink, J., and Kox, A.~J.~(2009). `On the history of the quantum. The HQ2 special issue', {\it Studies in History and Philosophy of Science} B: {\it Studies in History and Philosophy of Modern Physics}, 40, pp.~277-406.
\\
\\
Douglas, M.R., Kabat, D., Pouliot, P.,~and Shenker, S.H.~(1997). `D-branes and short distances in string theory', {\it Nuclear Physics B}, 485, pp.~85-127.
[hep-th/9608024].\\
\\
Douglas, M.R., Polchinksi, J., and Strominger, A.~(1997). `Probing five dimensional black holes with D-branes', {\it Journal of High Energy Physics}, 12, 003. 
[hep-th/9703031].\\ 
\\
Ehrenfest, P.~(1916). `Over adiabatische veranderingen van een stelsel in verband met de theorie der quanta', {\it Verslagen van de Koninklijke Akademie te Amsterdam}, 25, pp.~412-433; translation on pp.~79-93 in Waerden, B.L. van der (ed.), \emph{Sources of quantum mechanics}, New York: Dover, 1968.\\
\\
Ehrenfest, P.~(1923). `Le principe de correspondance', pp.~248-254 in: {\it Atomes et \'{e}lectrons. Rapports et discussions du conseil de physique tenu \`{a} Bruxelles du 1er au 6 avril 1921 sous les auspices de l'institut international de physique Solvay}. Paris: Gauthier-Villars.\\
\\
Ellis, G.~and Silk, J.~(2014). `Scientific method: Defend the integrity of physics',  \emph{Nature}, 516, pp.~321-323.
\\
\\
Ferrara, S., Kallosh, R., and Strominger, A.~(1995). `$N=2$ Extremal black holes', {\it Physical Review D}, 52, pp.~R5412-R5416. 
[hep-th/9508072].\\ 
\\
Ferrara,  S. and Kallosh, R.~(1996). `Supersymmetry and attractors',
 \emph{Physical Review D}, {54}, pp.~1514-1524.
  [hep-th/9602136].\\
\\
Fraser, D.~(2018). `The development of renormalization group methods for
 particle physics: Formal analogies between classical statistical mechanics and quantum field theory', \emph{Synthese}. doi.org/10.1007/s11229-018-1862-0.\\
\\
Gibbons, G.~and Townsend, P.K.~(1993). `Vacuum interpolation in supergravity via super p-branes', \emph{Physical Review Letters},  {71}, pp.~3754-3757.  
  [hep-th/9307049].\\
\\
Gopakumar, R., and Vafa, C.~(1999). `On the gauge theory/geometry correspondence', \emph{Advances in Theoretical and Mathematical Physics}, 5, pp.~1415-1443. 
[hep-th/9811131].\\
\\
Guica, M.~(2018). `An integrable Lorentz-breaking deformation of two-dimensional CFTs', \emph{SciPost Physics}, {5}, 048.
  [arXiv:1710.08415]. \\
  \\
Haco, S., Hawking, S.~W., Perry, M.~J., and Strominger, A.~(2018). `Black hole entropy and soft hair', {\it Journal of High-Energy Physics,}
1812, 098.
  [arXiv:1810.01847].\\
\\
Haghighat, B., Murthy, S., Vafa, C., and Vandoren, S.~(2016). `F-theory, spinning black holes and multi-string branches', {\it Journal of High-Energy Physics}, 1601, 009. 
[arXiv:1509.00455].\\
\\  
Hartman, T., Keller, C.A. and Stoica, B.~(2014). `Universal spectrum of 2d conformal field theory in the large $c$ limit', {\it Journal of High-Energy Physics}, {1409}, 118. 
  [arXiv:1405.5137].\\
\\
Hawking, S.W.~(1975). `Particle creation by black holes', \emph{Communications in Mathematical Physics},  {43}, pp.~199-220.
\\
\\
Horowitz, G.T., Maldacena, J.,~and Strominger, A.~(1996), `Non-extremal black hole microstates and U-duality', {\it Physics Letters B}, 383, pp.~151-159. 
[hep-th/9603109]. \\
\\
Horowitz, G.T.,~and Polchinski, J.~(1997). `A correspondence principle for black holes and strings', {\it Physical Review D}, 55, pp.~6189-6197. 
[hep-th/9612146].\\
\\
Huggett, N.~(2017). `Target space $\not=$ space', {\it Studies in History and Philosophy of Science} B: {\it Studies in History and Philosophy of Modern Physics}, 59, pp.~81-88. 
[arXiv:1509.06229].\\
\\
Humphreys, P.~(2016). {\it Emergence. A Philosophical Account}. Oxford: Oxford University Press.\\
\\
Jammer, M.~(1989). {\it The Conceptual Development of Quantum Mechanics}. Los Angeles, CA: Tomash. \\
\\
Joas, C. and Lehner, C.~(2009). `The classical roots of wave mechanics: Schr\"{o}dinger's transformations of the optical-mechanical analogy', {\it Studies in History and Philosophy of Science} B: {\it Studies in History and Philosophy of Modern Physics}, 40, pp.~338-351.\\
\\
Johnson, C.~V.~(2003). {\it D-Branes}. Cambridge: Cambridge University Press.\\
\\
Katz, S.H., Klemm, A., and Vafa, C.~(1999). `M theory, topological strings and spinning black holes', {\it Advances in Theoretical and Mathematical Physics}, 5, pp.~1445-1537. 
[hep-th/9910181].\\
\\
Katzir, S., Lehner, C., and Renn, J.~(2013). {\it Traditions and Transformations in the History of Quantum Physics, HQ3}. Berlin: Edition Open Access.\\
\\
Kawai, H., Lewellen, D.C., and Tye, S.H.~(1986). `A relation between tree amplitudes of closed and open strings', {\it Nuclear Physics B}, 269, pp.~1-23.
   \\
\\
Khoury, J. and Verlinde, H.~(2000).
`On open/closed string duality', 
\emph{Advances in Theoretical and Mathematical Physics},  {6}, pp.~1893-1908. 
  [hep-th/0001056].\\
\\  
Kiem, Y., Verlinde, H., and Verlinde, E.~(1995). `Black hole horizons and complementarity', {\it Physical Review D}, 52, pp.~7053-7065. 
[hep-th/9502074].\\
\\  
Linnemann, N.S. and Visser, M.R.~(2018). `Hints towards the emergent nature of gravity', {\it Studies in History and Philosophy of Science} B: {\it Studies in History and Philosophy of Modern Physics}, 64, pp.~1-13. 
[arXiv:1711.10503]. \\
\\
 Maldacena,  J.M.~(1998a). `Branes probing black holes',
  \emph{Nuclear Physics B. Proceedings Supplements},  {68}, pp.~17-27. 
  [hep-th/9709099]. \\
 \\ 
Maldacena, J.M.~(1998b). `The large $N$ limit of superconformal field theories and supergravity', {\it Advances in Theoretical and Mathematical Physics}, 2, pp.~231-252. 
  [hep-th/9711200].\\
\\
Maldacena, J.~M. and Strominger, A.~(1997a). `Black hole grey body factors and D-brane spectroscopy', \emph{Physical Review D}, 55, pp.~861-870.  
[hep-th/9609026]. \\
\\
McGough,  L., Mezei, M.,~and Verlinde, H.~(2018). `Moving the CFT into the bulk with $ T\overline{T} $', {\it Journal of High-Energy Physics}, {1804}, 010.
[arXiv:1611.03470].\\
\\  
Nagel, E.~(1961). {\it The Structure of Science. Problems in the Logic of Scientific Explanation}. New York: Harcourt.\\
\\
Navarro, J., Blum, A., and Lehner, C.~(2017). `On the history of the quantum, HQ4', {\it Studies in History and Philosophy of Science} B: {\it Studies in History and Philosophy of Modern Physics}, 60, pp.~1-148. 
\\
\\
O'Connor, T.~and Wong, H.Y.~(2015). `Emergent properties'. {\it The Stanford Encyclopedia of Philosophy}. https://plato.stanford.edu/entries/properties-emergent.\\
\\
Ooguri, H., Strominger, A., and Vafa, C.~(2004). `Black hole attractors and the topological string', {\it Physical Review D}, 70, 106007.
[hep-th/0405146].\\
\\
Polchinski, J.~(1995). `Dirichlet branes and Ramond-Ramond charges', {\it Physical Review Letters}, 75, pp.~4724-4727. 
[hep-th/9510017].\\
\\
Radder, H.~(1991). `Heuristics and the generalized correspondence principle', {\it British Journal for the Philosophy of Science}, 42, pp.~195-226. 
\\
\\
Read, J.~and M\o ller-Nielsen, T.~(2018). `Motivating Dualities'. {\it Synthese},  doi:10.1007/s11229-018-1817-5;  http://philsci-archive.pitt.edu/14663.\\
\\
Rovelli,  C. (2013). `A critical look at strings', \emph{Foundations of Physics}, 43, pp. 8-20.  
\\
\\
Rynasiewicz, R.~(2015). `The (?) correspondence principle', pp.~175-199 in  Aaserud, F. and Kragh, H. (eds.), \emph{One hundred years of the Bohr atom}, Copenhagen: The Royal Danish Academy of Sciences and Letters.\\
\\
Smirnov, F.A. and Zamolodchikov, A.B.~(2017). `On space of integrable quantum field theories', \emph{Nuclear Physics B}, {915}, pp.~363-383.
  [arXiv:1608.05499].\\
\\
Smolin, L.~(2006). {\it The Trouble with Physics. The Rise of String Theory, the Fall of a Science, and What Comes Next}. Boston: Houghton Mifflin. \\
\\
Strominger, A.~(1996).
  `Macroscopic entropy of N=2 extremal black holes', \emph{Physical Letters B}, {383}, pp.~39-43.
  [hep-th/9602111].\\
\\
Strominger, A.~and Vafa, C.~(1996). `Microscopic origin of the Bekenstein-Hawking entropy',  {\it Physics Letters B}, 379, pp.~99-104.   
  [hep-th/9601029].\\
\\  
Susskind, L.~(1998). `Some speculations about black Hole entropy in string theory', pp.~118-131 in: Teitelboim, C.~and Zanelli, J.~(eds.), {\it The Black Hole, 25 Years After:} Proceedings (17-21 Jan 1994, Santiago, Chile).  Singapore: Wold Scientific. [hep-th/9309145].\\
\\
Susskind, L., Thorlacius, L., and Uglum, J.~(1993). `The stretched horizon and black hole complementarity', {\it Physical Review D}, 48, pp.~3743-3761. 
[hep-th/9306069].\\
\\  
Townsend, P.K.~(1995). `The eleven-dimensional supermembrane revisited', {\it Physics Letters B}, 350, pp. 384-389. 
[hep-th/9501068].\\
\\
Townsend, P.K.~(1996). `D-branes from M-branes',  {\it Physics Letters B}, 373, pp.~68-75. 
[hep-th/9512062].\\
\\
Vafa, C.~(1996). `Gas of $D$-branes and Hagedorn density of BPS states', {\it Nuclear Physics B}, {463}, pp.~415-419.
  [hep-th/9511088].\\
\\  
Vafa, C.~(1998). `Black holes and Calabi-Yau threefolds', {\it Advances in Theoretical and Mathematical Physics}, 2, pp.~207-218.
  [hep-th/9711067].\\
\\
Verlinde, E.~(2011). `On the origin of gravity and the laws of Newton', {\it Journal of High Energy Physics}, {1104}, 029. 
  [arXiv:1001.0785].\\
\\
Wit, B. de, Hoppe, J., and Nicolai, H.~(1988). `On the quantum mechanics of supermembranes', {\it Nuclear Physics B}, 305, pp. 545-581. 
\\
\\
Wit, B. de, Luscher, M., and Nicolai, H.~(1989). `The supermembrane is unstable', {\it Nuclear Physics B}, 320, pp. 135-159. 
\\
\\
Witten, E.~(1995). `String theory dynamics in various dimensions', {\it Nuclear Physics B}, 443, pp.~85-126.
  [hep-th/9503124].\\
\\
Witten, E.~(1996a). `Bound states of strings and p-branes', {\it Nuclear Physics B}, 460, pp.~335-350. 
  [hep-th/9510135].\\
\\
Witten, E.~(1996b). `Five-branes and $M$-theory on an orbifold', {\it Nuclear Physics B}, 463, pp.~383-397.
  [hep-th/9512219].\\
\\
Wong, H.Y.~(2010). `The secret lives of emergents', pp. 7-24 in: Corradini, A.~and O'Connor, T. (eds.), {\it Emergence in Science and Philosophy}.  New York and London: Routledge.\\

\end{document}